\documentclass[aps,pre,twocolumn, citeautoscript,superscriptaddress]{revtex4-1}
\usepackage{color}
\usepackage{graphicx}
\usepackage[margin=1in]{geometry}
\usepackage{amsfonts, amssymb,amsmath, dsfont,MnSymbol}
\usepackage{relsize}%
\usepackage{textpos}
\usepackage{float}
\usepackage[colorlinks=true,linkcolor=blue]{hyperref}%

\newcommand{\ud}{\mathrm{d}}

\begin{document}
\title{Understanding and modeling polymers: The challenge of multiple scales}
\date{}
\author{Friederike Schmid}
\email{friederike.schmid@uni-mainz.de}
\affiliation{Institut f\"{u}r Physik, Johannes Gutenberg-Universit\"{a}t Mainz, Staudingerweg 9, 55128 Mainz, Germany}


\begin{abstract}

Polymer materials have the characteristic feature that they are
multiscale systems by definition. Already the description of a single
molecules involves a multitude of different scales, and cooperative
processes in polymer assemblies are governed by the interplay of these
scales. Polymers have been among the first materials for which
systematic multiscale techniques were developed, yet they continue to
present extraordinary challenges for modellers. In this perspective,
we review popular models that are used to describe polymers on
different scales and discuss scale bridging strategies such as static
and dynamic coarse-graining methods and multiresolution approaches. We
close with a list of hard problems which still need to be solved in
order to gain a comprehensive quantitative understanding of polymer
systems on all scales.

\end{abstract}

\maketitle

%
%

\section{Introduction} 
\label{sec:intro}

Multiscale problems are omnipresent in materials science.  The
properties of most materials typically result from a combination of
many processes on vastly different length and time scales, ranging from
electronic excitations and atomic or molecular vibrations on the
Angstrom and femtosecond scale to material fatigue on time scales over
several years. In polymeric systems, disentangling the different
characteristic scales scales that determine their behavior is
particularly difficult. This is because the relevant molecular length
scales -- which range from the scale of the local chemical monomer
structure to the scale of chain conformations -- strongly overlap
with the relevant length scales of the next level of inter-molecular
and possibly supra-molecular organization, and these in turn overlap
with the length scales of continuum mechanics on which  materials are
described in terms of elastohydrodynamic equations.  Therefore, an
all-inclusive, comprehensive modeling of a polymeric
system \cite{GJ_19} remains a formidable challenge despite decades of
theoretical efforts \cite{BBDG_00,MPlathe_02,Review_Peter_10,
Review_Noid_13, Review_Liu_13,Review_vanderVegt_13,Review_Voth_13,
Review_SA_14, Review_BG_16, Review_Rudzinski_19, Review_SPH_19,
Review_Memory_21,Review_Webb_21, NTL_22, Review_Schilling_22}. 


In the present perspective article, we discuss some selected
approaches to this problem, focussing on recent developments.  Before
doing so, we will quote a few examples of scale-bridging phenomena in
polymers that inherently require multiscale descriptions.  

The first and most basic example is the emerging viscoelasticity and
viscoplasticity in {\em polymer rheology}, a field where the
multiscale character of polymer-based systems is immediately
apparent \cite{doi_edwards_book}. Polymeric materials respond to
applied stress with some time delay (memory), a clear signature of an
incomplete separation of time scales. This is because the time scales
of intramolecular (internal chain) relaxation cannot be separated
from the time scales of diffusion and intermolecular re-organization,
in particular in the presence of entanglements \cite{Review_Liu_13}.

Another prominent classical multiscale phenomenon in polymer science is
{\em polymer crystallization} \cite{handbook_polymer_crystallization_13},
which involves local crystallization on the monomer scale, the
formation of crystalline lamellae on the mesoscale, and the macroscale
organization of lamellae, often into spherulites \cite{CS_16}. Already
the local structure is not necessarily unique \cite{Beran_16}, but may
result from a competition of several polymorphs depending on the
processing \cite{GVDP_90}.  Predicting such polymorphs requires accurate
theoretical descriptions at the electronic structure level \cite{CS_16}
as well as multiscale modelling approaches to enable studying the kinetics of
self-assembly \cite{LKB_18,LBVK_20}. On the mesoscale, the mechanisms
that determine and eventually constrain the growth of crystalline
lamellae are still under debate \cite{SSST_22}. One particular
intriguing phenomenon is the ''melt memory'' effect \cite{SCM_20,
Muthu_16}: Even after melting a polymer crystal, the melt retains some
knowledge about the previous structure and tends to re-crystallize at
previously crystalline positions after cooling.  Recent systematic
simulation studies by Luo, Sommer and others have suggested that the
thickness of crystalline lamellae is determined by the entanglement
length in the melt prior to crystallization \cite{LS_14, LKS_16,
LKS_17, XLYS_17, ZFMP_19, PS_21} -- consistent
with the experimental observation that the time scale on which the
melt memory survives roughly matches the time scale of re-entanglement
kinetics \cite{LY_20}. Such findings illustrate how mesoscale
structure (entanglements) can have a profound effect on local
structure (local packing), and vice versa, in polymers.  The
associated time scales can be very large, which offers unusually
versatile opportunities to control local structure by
processing \cite{CBCF_19}, for example, in flow-induced crystallization
\cite{HPSK_20, CMTS_20, LWGH_21, KSCM_21, SCCL_22}, or by tiny
chemical modifications \cite{HPSK_21,FMC_21}.  On the other hand, the
mesoscale structure and dynamics determines the elastic and plastic
response of the materials to deformations \cite{JRLM_15,JLPR_17} and
the inhomogeneous stress fields in the materials, which in term drive
the large-scale structure formation and spherulite
growth \cite{Schultz_12}.

The interplay of multiple scales also determines the structural and
dynamic properties of other {\em multiphase polymer materials}
\cite{handbook_multiphase_polymer_11} that are highly heterogeneous
and filled by internal interfaces, such as polymer blends
\cite{SA_09,XZH_12}, block copolymer melts and solutions
\cite{CH_04,OW_11,ME_12,KKP_22}, or foams \cite{KPY_21}. It is
particularly prominent in {\em polymer nanocomposites}
\cite{Review_BER_06, Review_ZYL_08, Review_RAFP_17, KCWK_19, GCJK_21},
where fillers are introduced, e.g., to improve the mechanical
properties of a polymeric matrix. The molecular origins of the
resulting mechanical reinforcement are diverse, they include a
redistribution of strain in the polymer matrix \cite{SMSF_21} as well
as stretching of chains at the interfaces \cite{CBBX_16}. A detailed
knowledge of the structure of the material on both local and
mesoscopic scales is thus necessary to understand the macroscopic
viscoelastic properties of the materials. Likewise, transport
properties such as the thermal conductivity \cite{ZZXZ_18} depend on
the microscopic structure in the bulk matrix as well as at interfaces,
i.e., the Kapitza resistance \cite{SYCY_13,HMZR_16}, and on the
mesoscale shape and spatial distribution of the fillers.

Finally, {\em biomaterials} provide some of the most sophisticated
polymer-based multiscale materials, due to their characteristic
hierarchical structure. A well-known example is is spider silk
\cite{ESS_11, BYTM_19}, which also serves as a good example how the
properties of such materials crucially depend on the way how they has
been processed (in this case spun) \cite{JK_03}.  Other examples are
fibers made of collagen, which are abundant in mammals
\cite{KG_02,Buehler_06, SR_09,Bella_16}.  Collagen is found in the
extracellular matrix of tissues as different as skin, fascia,
cartilage and bones, and is to a great extent responsible for their
superior material properties. Twenty-nine types of collagen
have been reported in the literature \cite{KG_02}, the most frequent
being collagen I, which is present in, e.g.,  dermis, tendon, and
bone.  The primary structure of collagen peptides is characterized by
repeats of three residues Gly-X-Y, e.g., Gly-Pro-Hyp. The quaternary
structure is a triple helix, where three polypeptide strands wrap
around each other to form a helix of length \mbox{$\sim 300$ nm}, the
tropocollagen. Staggered arrays of tropocollagen self-assemble
(spontaneously \cite{KABP_06}) to fibrils, the building blocks of
fibers (with size roughly \mbox{10 $\mu$m}), which then aggregate to even
larger structures \cite{OIMW_06}. Interestingly, the collagen triple
helix seems to be only marginally stable, it melts at temperatures
just slightly above, or even below the body temperature
\cite{LMKL_02}. This suggests, on the one hand, that the triple helix
is stabilized by the fibrillar/microfibrillar suprastructure
\cite{KABP_06}, but also, on the other hand, that collagen frequently
unfolds and refolds on a local scale. The combination of strength and
softness would then contribute to the unique material properties of
collagen tissues, to their elasticity, and to their capacity to
dissipate sudden energy bursts. 

In the context of the {\em living tissues}, protein fibers are only
one building block in the even more complex multiscale structures of,
e.g., bones \cite{BSG_16, GNHP_21} or skin \cite{ALS_19}. One
important aspect of living materials is their dissipative character:
They are kept alive by constant energy consumption and thus never
reach thermal equilibrium, nor a thermally metastable state, but
continuously produce entropy. Prominent representatives of such
inherently nonequilibrium materials are protein filament structures
which form the cytoskeleton of cells and are responsible for their
mechanical elasticity as well as their motion and/or contraction
\cite{Shelley_16, PMRD_21}.  Another example is the recently
discovered phenomenon of liquid-liquid phase separation (LLPS) in
cells \cite{HWJ_14,SB_17, BAFM_18, BBH_18, PHP_18, AD_19, CHP_20,
LPR_21}: Certain proteins mediate the formation of nanosized
condensates in cells -- so-called membraneless organelles -- which
helps to organize cellular content and possibly contributes to gene
regulation. Whereas the phase separation itself is driven by
thermodynamic interactions, the size and location of the droplets is
most likely controlled by nonequilibrium, energy-consuming processes. 

These selected examples illustrate the omnipresence of multiscale
phenomena in polymeric systems, in seemingly simple ones such as
one-component polymer melts as well as in  complex ones such as
functional polymers in a nonequilibrium living matter context. This
multiscale character of polymers presents an outstanding challenge for
modellers.  

Synthetic polymer systems have been among the materials for which
systematic multiscale modeling methods have been developed, which 
related particle-based coarse-grained models to real polymers
such as polyethylene \cite{BBPL_91,RM_96, TPBB_97, CM_97, DM_97,
DM_98}, polycarbonates \cite{PBKH_91, PP_94, TKBB_98,TKHB_98,
EZTH_99}, and others \cite{DRM_96,HM_98}. These early studies already
addressed key challenges that are still subject of active research
today: (i) The coarse-graining procedure, i.e.  constructing
coarse-grained models using input from quantum chemical calculations
and/or atomistic models \cite{PBKH_91, BBPL_91, PP_94,RM_96,
DRM_96,CM_97,HM_98,TPBB_97, TKBB_98}; (ii) Reverse backmapping, i.e.,
the reconstruction of an atomistic configuration from a coarse-grained
configuration \cite{DM_97, TKHB_98, EZTH_99}; (iii) Dynamic mapping
\cite{DM_98,TKHB_98}, i.e., the question how to extract dynamical
information from the coarse-grained simulations. 

Since then, much progress has been made in the field of multiscale
modelling of polymers and of soft matter systems in general, and
several excellent reviews have highlighted different aspects of the
problem, see, e.g., Refs.\cite{Review_Peter_10, Review_PB_11,
Review_Noid_13, Review_Liu_13, Review_vanderVegt_13, Review_Voth_13,
Review_BG_16, Review_Rudzinski_19, Review_NTMC_20,
Review_Memory_21,Review_Webb_21, NTL_22}. Nevertheless, central
challenges still persist. In the present perspective article, we 
discuss the current situation in the light of state of the art and
recent progress. We begin with a rough outline of models that are
used to describe polymeric systems on different scales. Then we 
discuss a number of scale-bridging strategies that have been developed
in the past and have been used for polymeric systems or might be
applicable for them. We close with an outlook on open problems
for the future.

\section{Scales in Polymers}
\label{sec:scales}

To set the stage, we begin with discussing the different
scales that are involved in our multiscale picture of polymers,
and introduce classes of polymer models that are used to study
polymer materials on these different scales.

\subsection{Monomer/Oligomer scale: The scales of chemical specifity}
\label{sec:scales_monomers-oligomers}

The basic building blocks are the {\em monomers}. They can have a simple
chemical structure, as in the case of many commodity polymers such as
polystyrene, or a rather complicated structure, as in the case of
biopolymers such as RNA, DNA, or proteins. The structure of the monomers
on the monomer scale determines local properties such as the charges 
and the polarization, the solubility in a solvent \cite{barton_book}, 
the existence and structure of a hydration shell \cite{HH_98}, the local
affinity to surfaces \cite{HR_16}, or -- in studies of polymer reactions,
the monomer reactivity \cite{BM_08}. In general, these properties are also
influenced by the larger scale structure of polymer systems. For
example, the effective monomer reactivity depends on the accessibility
of the reactive sites, which is determined not only by the local
electronic and steric monomer structure, but also by the polymer
conformation \cite{BM_08,VCVR_21}. Likewise, the effective charges and/or
polarization of monomers depend on the local environment \cite{RG_91,
WKS_12}. In most cases, however, the corrections due to the larger scale
structure are small compared to the intrinsic value imposed by the
monomer structure. To study polymers on the monomer scale, atomistic
models are used, and in some cases quantum mechanical
modelling is necessary \cite{WU_06,DEOZ_18}.

The next level of organization is the {\em oligomer} scale, i.e., the
scale of short polymer sections and monomer-monomer interactions.  On
that scale, cooperativity effects due to non-bonded or bonded
interactions between monomers start to become prominent and even
dominate. Here, the term ''non-bonded'' refers to general interactions
between monomers of given types, no matter whether or not they belong
to the same molecule (e.g., electrostatic interactions or van der
Waals interactions), and the ''bonded'' interactions subsume the
additional interactions between monomers that are close neighbors in
the molecule (e.g., chemical bonds, bending or torsional potentials).
Emerging properties of interest on the oligomer scale are, e.g., the
effective monomer-monomer incompatibility
\cite{polymer_compatibility_book_82}, ion-specific effects
\cite{BKNN_05, MMK_14}, the propensity to crystallize
\cite{handbook_polymer_crystallization_13}, or solvency/cosolvency
effects \cite{MMK_14,LWMJ_22}.  Again, these properties also depend on
the higher order organization, e.g., as has been discussed in the
introduction for the case of crystallization.  The modelling at this
level is still often based on atomistic force fields, but chemically
specific force fields such as the celebrated MARTINI
model \cite{MRYT_07} are starting to become useful, see also the recent
review by Dhamankar and Webb \cite{Review_Webb_21}.  In such
coarse-grained models, several atoms are lumped into one effective
particle, and the (bonded and non-bonded) interactions between
particles are determined either in a bottom-up fashion from atomistic
simulations, or in a top-down fashion from experimentally accessible
data, or from a combination of the two. For an overview over
coarse-graining approaches, we refer to the excellent review of W.\
Noid \cite{Review_Noid_13} (see also below in Sec.\
\ref{sec:multiscale_static_structure-based}).

\subsection{Polymer scale: The scale of conformations}
\label{sec:scales_polymers}

The third level, the {\em polymer} scale, is the realm of classical
polymer physics, where generic statistical mechanics approaches have
celebrated successes \cite{degennes_book, doi_edwards_book,
rubinstein_book}.  At this level, scaling laws have scored victories,
both regarding static and dynamic properties of polymeric systems, and
simple calculations based on ''scaling blobs'' \cite{Halperin_94,
rubinstein_book} can make meaningful predictions. This is because
polymer molecules consisting of many identical monomer units start to
exhibit universal behavior beyond a certain molecular weight.
Therefore, renormalization groups concepts can be applied, according
to which the fractal large-scale structure of polymer conformations
does not depend on details of the local monomer structure. This
results in the paradigm of the ''Gaussian chain model''
\cite{doi_edwards_book}, which describes a polymer molecule as a
random walk in space.  In the case of complex heteropolymer molecules
such as intrinsically disordered peptides (IDPs), applying scaling
concepts is more challenging, but still at least partially successful
\cite{DP_13,PPLM_14,SKBH_14}.  Theoretical models at this level are
mostly based on effective phenomenological parameters
\cite{doi_edwards_book} such as the Kuhn length, the famous Flory
Huggins $\chi$-parameters characterizing polymer-solvent or
monomer-monomer interactions, the monomer mobility, the effective
monomer charge and possibly the Debye screening length.  

A number of generic coarse-grained simulation models have been
proposed already decades ago to study polymer properties at this
level: Lattice models, where polymers are modeled as random
self-avoiding walks on a lattice, and off-lattice models, where
polymers are modelled as chains of interacting hard-core spheres
connected by springs. Among the most prominent models of this type is
the Bond Fluctuation Model \cite{CK_88}, a lattice model where
monomers occupy cubes on a lattice and can be connected by a finite
set of bonds, and the Kremer-Grest model \cite{GK_86}, an off-lattice
model that represents polymers as strings of hard-core spheres
connected by nonlinear springs. Such models can be extended in various
ways, e.g., to include bending potentials \cite{FKA_99, Hsu_14},
attractive non-bonded interactions \cite{DB_91}, or (in the case
polymer solutions) a hydrodynamic coupling to a fluid medium
\cite{AD_99}.  They have been used to verify scaling predictions
\cite{PBHK_91, Hsu_14}, to study generic aspects of single polymer
phase transitions such as chain adsorption or the coil-globule
transition \cite{VLFI_02, BBMP_06, LRPB_08} properties of polymer
melts and blends \cite{Mueller_99} and even dynamical transitions such
as the glass transition \cite{RBB_93,BBP_03,RPM_21}. To some extent,
they can also be used to make quantitative predictions for specific
polymers.  For example, recent work by Everaers and coworkers
\cite{EKHF_20} has shown that, for a wide range of commodity polymer
melts, matching a single local property in melts of Kremer-Grest
chains -- the so-called dimensionless Kuhn number -- is sufficient to
reproduce the correct entanglement modulus \cite{ESGS_04,EKHF_20}. The
Kuhn number is derived from microscopic quantities, i.e., the number
of Kuhn segments in a volume of Kuhn length cube. The entanglement
modulus is roughly proportional to a macroscopic quantity, the plateau
shear modulus.  Hence this example shows how simulations of a properly
matched generic polymer model can be used to predict important
characteristics of polymer materials.  Milner has recently proposed a
universal scaling theory of entanglements for melts and solutions of
chains with arbitrary flexibility \cite{Milner_20}, suggesting that
similar approaches might also be successful for polymer solutions.  


\subsection{Interacting polymers: The blob scale}
\label{sec:scales_blobs}

The properties of {\em polymer systems} containing many polymers are
often determined by conformational restructuring on scales that are
much larger than the monomeric scale. On such scales, polymers behave
in many respect like single soft, interpenetrating ''blobs'', or
chains of such blobs.  In polymer physics, the term blob often refers
to a theoretical framework that allows for simple intuitive
derivations of crossover phenomena between different scaling regimes
in polymer solutions \cite{Halperin_94,rubinstein_book}. Here we will use it more generally to
describe the soft character of overlapping polymers.

In large-scale studies of interacting polymers, two novel classes of
simulation models are becoming increasingly popular that account for
this soft character: Ultra-coarse-grained particle-based models with
soft potentials, and density-based models. In soft potential models,
coarse-grained units are assumed to represent lumps of a sufficiently
large number of microscopic particles that they can interpenetrate
each other.  The non-bonded potentials are still described in terms of
pair interactions between particles with positions $\vec{r}_i$ and
$\vec{r}_j$
\begin{equation}
\label{eq:potential_pair}
U_{\mbox{\footnotesize nb}} [\vec{r}_k]
  = \sum_{i,j} V_{ij}(\vec{r}_i,\vec{r}_j),
\end{equation}
possibly augmented by higher-order multibody potentials \cite{PF_01}.
However, the potentials do not diverge at $\vec{r}_i=\vec{r}_j$.  This
is the case, for instance for ''dissipative particle dynamics''
(DPD)-models \cite{DPD_97, PF_01} or blob models \cite{LOS_91, MK_98,
Likos_01, BLHM_01, EM_01, VBK_10, NLMC_14, AMPP_15}. 

In contrast, in density-based models \cite{LZ_94, BGZ_00, GP_03,
DKDM_08, MK_09, Wang_09, Mueller_11, ZMD_16, BC_21}, the non-bonded
potentials are expressed as a functional of local number densities
$\underline{\rho} = \{\rho_\alpha\}$ of coarse-grained monomer or
solvent particles of type $\alpha$, typically in the form of an
integral over a ''free energy density''
\begin{equation}
\label{eq:potential_density}
U_{\mbox{\footnotesize nb}}[\underline{\rho}] 
=  \int \ud^3 r \: f(\vec{r}, \underline{\rho}).  
\end{equation}
The function $f(\vec{r},\underline[\rho])$ is often taken to have a local
quadratic form, defined in terms of Flory Huggins-like interaction
parameters
\begin{equation}
\label{eq:f_rho}
\frac{f(\vec{r},\underline{\rho}) }{k_B T}
= \sum_{\alpha \beta} \chi_{\alpha \beta} 
  \rho_\alpha(\vec{r}) \rho_\beta(\vec{r})
+ \frac{\kappa}{2} (\sum_\alpha \rho_\alpha(\vec{r}) - \rho_0)^2,
\end{equation}
where $\kappa$ gives the compressibility of the polymer solution or
melt, and  we have assumed for simplicity that the volume fraction per
bead of all entities is the same We should mention that in practical
simulations, the value of $\kappa$ is often reduced quite
substantially compared to the real compressibility of polymers.  This
is done to to avoid numerical instabilities and enable simulations
with larger time steps.

To complete the definition of the model, one must also specify how to
determine the local densities $\rho_{\alpha}$ and how to formulate the
corresponding spatially discretized version of the equations of
motion.  Often, the local densities are evaluated on a grid
\cite{LZ_94}, but other off-lattice variants based on weighted
densities have also been proposed \cite{HM_10, Mueller_11}. When using
a grid-based model in dynamical simulations, a second practical issue
is how to determine the resulting forces on monomers -- whether to
directly take the derivative of the discretized Hamiltonian with
respect to the monomer positions \cite{ZQKS_17,ZQKS_18},  or whether
to calculate a discretized force field and interpolating that
\cite{MK_09,SSKM_17, BC_21}.  The former strategy guarantees that the
simulation is based on a well-defined Hamiltonian, but it introduces
lattice artefacts. The latter strategy gives more freedom to reduce
the lattice artefacts and (approximately) restore momentum
conservation in molecular dynamics simulations, but it does not
guarantee that one samples a rigorously defined statistical ensemble
in the limit of zero time step. Thus the former approach is better
suited for studying the statistical mechanics of the system, and the
latter for studying processes where hydrodynamics is important. 

Eq.~(\ref{eq:f_rho}) defines one of the simplest density-based models,
but numerous extensions are possible to make the model more flexible.
One can add higher order terms \cite{Mueller_11, ZMD_16}, additional
density fields that characterize, e.g., local orientation or charges
\cite{DRK_12} and/or nonlocal terms.  For example, electrostatic
interactions can be included in (\ref{eq:potential_density}) by
including the energy density of the electrostatic field generated by
the charge density distribution $\underline{\rho}(\vec{r}')$.

Both soft-potential models and models with density-based potentials are
particle-based and describe the polymers as connected chains of explicit
monomers, which differ only in the type of non-bonded interactions.
Removing the hard excluded-volume interactions, however, has a
fundamental consequence: It removes topological interactions, i.e., the
chains can now cross each other. This significantly changes the dynamic
properties of the coarse-grained models and -- in some cases -- even the
static structure. 

Most prominently, the conformations of strictly two-dimensional polymers
in dense melt are radically different for overlapping and
non-overlapping polymers \cite{DS_86, DS_87, SJ_03, CMB_05,CMWJB_05,
MWKJ_10}. The configurations of overlapping polymers are rather open and
the number of interchain contacts per monomer is roughly constant.  In
contrast, non-overlapping polymers segregate from one another, and the
number of contacts per monomer scales as $N^{-3/8}$ with increasing
chain length $N$.  This is because most open configurations are
forbidden due to excluded-volume interactions. This effect is
independent of dynamics and also persists in Monte Carlo simulations
that simply sample the phase space.

In higher dimensions, the fraction of actually forbidden conformations
in phase space is negligible, and the effect of hard excluded-volume
interactions is more subtle. In systems of closed (ring) polymers,
topological constraints partition the phase space, since a large set
of energetically allowed conformations cannot be accessed kinetically
from a given start configuration:  Initally concatenated rings cannot
be separated and initially separated rings cannot be concatenated.  As
a result, ring polymers in a melt of non-concatenated polymers are
more compact than linear polymers, and their size (radius of gyration)
scales differently as a function of $N$ \cite{HLGG_11}.  Capturing
such effects with soft coarse-grained models is a formidable
challenge.  Narros {\em et al} \cite{NLMC_14} have proposed a
hierarchical multi-blob approach, where the direct interactions
between soft blobs (coarse-grained monomers) are supplemented by
additional interactions between the centers of mass that account for
the effect of topological interactions in a statistical sense. With
this approach, they could reproduce the shrinking of ring polymers in
melts with the correct exponent.  However, other characteristics of
large ring polymers in ring polymer melt, e.g., the dominance of
double-folded conformations with primitive tree structure (''lattice
animals'') \cite{RE_14,EGRR_17}, are not captured. Interestingly, a
recent comparison of density functional calculations (which ignore
topological constraints) and and CG simulations of Kremer-Grest chains
has suggested that topological effects have no effect on the density
profiles in sufficiently dense melts \cite{CLE_21}, although they do
seem to affect the thickness of depletion regions close to surfaces in
semidilute solutions.

In contrast to ring polymer melts, melts of linear polymers are
ergodic in phase space and blob models can mostly account for their
static structure, at least on large scales. On small scales, there are
deviations. For example, they tend to overestimate the frequency of
small knots \cite{ZMVD_20, WASM_21}, in particular if the size of the
excluded volume of monomers is comparable to that of the Kuhn segment
\cite{ZMVD_20}.  More importantly, models with soft interactions fail
to reproduce their dynamics at large $N$ which is characterized by
entanglements between polymers as already discussed earlier
\cite{doi_edwards_book, Review_Masubuchi_14, Review_SA_14}: According
to the classic reptation picture, polymers undergo an effective
one-dimensional diffusion in a tube, which is created by their
entanglements with other polymers.

\begin{figure}[t]
\includegraphics[width=0.35\textwidth]{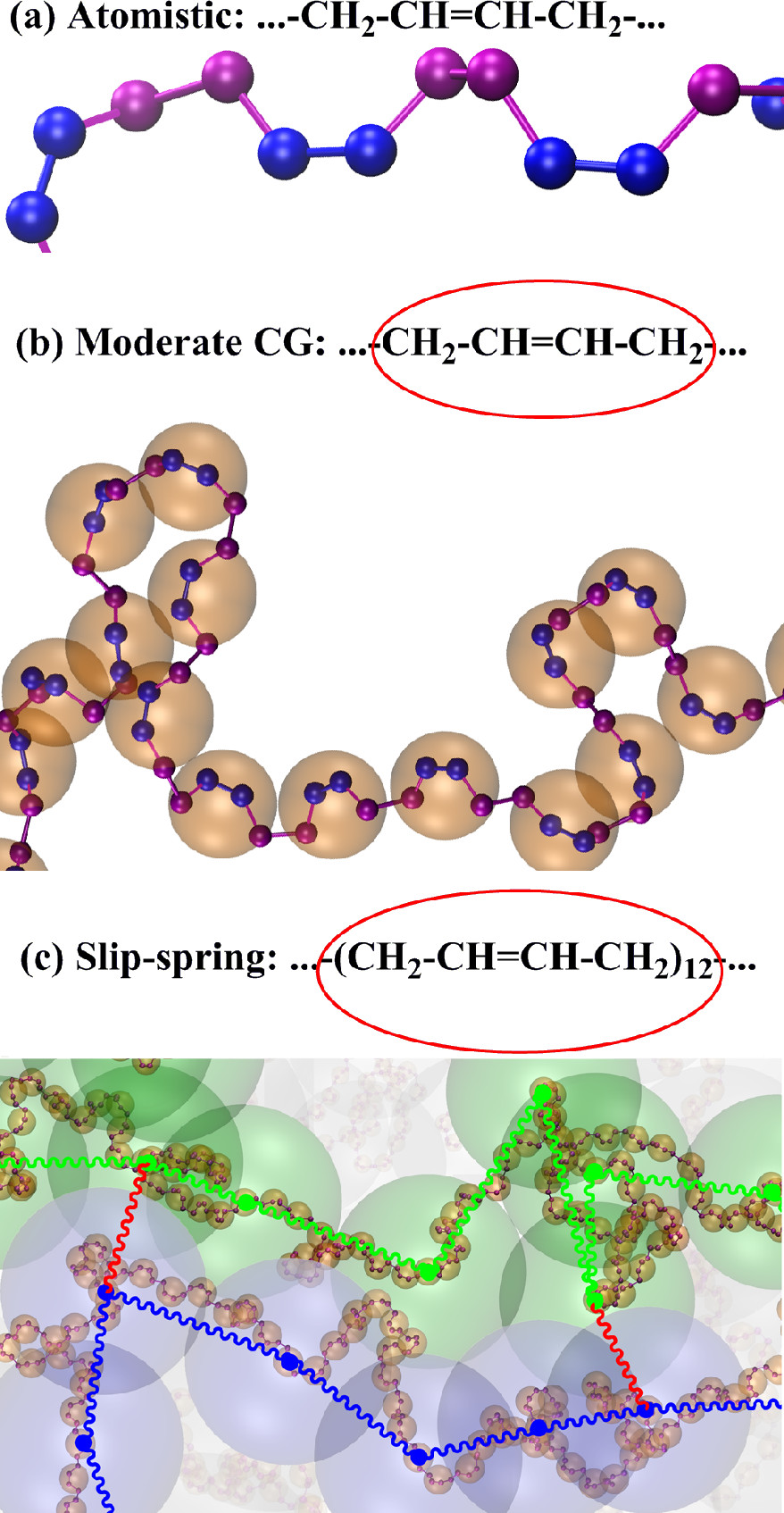}
\caption{Different levels of description of 
{\em cis}-1,4-polybutadiene (cPB) in Ref.\ \cite{BSRC_21}. 
(a) United atom model;
(b) Structurally coarse-grained model with hard core interactions.
(c) Soft potential model with slip-links. The dynamical single-
chain and viscoelastic properties can be mapped onto each other and
are also in good agreement with experimental data. 
Reprinted from Ref.\ \protect\cite{BSRC_21}  with
permission of XXX}
\label{fig:BSRC_21}
\end{figure}

Polymers interacting with soft potentials, however, do not reptate,
\cite{PB_01}.  Schieber \cite{Schieber_03} and later Likhtman
\cite{Likhtman_05} have proposed an ingenious way to restore
entanglements at the level of single-chain dynamics \cite{MD_08,
SAC_21, Review_SA_14}: They proposed to mimic the effect of
entanglements by virtual ''slip links'', discrete objects through
which the chains must slip. This model has later been extended to
multi-chain models where the slip links are fluctuating objects that
connect different chains to each other
\cite{UM_12,CMZM_12,RPAS_15,RPSA_17}.  Such slip link degrees of
freedom introduce effective attractive interactions between polymers.
However, they can be calculated analytically and subtracted from the
basic potential function, e.g., Eqs.\ (\ref{eq:potential_pair}) or
(\ref{eq:potential_density}), to eliminate their effect on the static
behavior \cite{UM_12,CMZM_12}. Wu et al and Behbahani et al have
recently demonstrated the potential of multi-chain slip-link
approaches in ultra-coarse-grained simulations of real commodity
polymers such as polyethylene \cite{WKNM_21}, polystyrene
\cite{WMM_21}, and polybutadiene \cite{BSRC_21} (see Fig.\
\ref{fig:BSRC_21}).

\begin{figure*}[t]
\includegraphics[width=0.9\textwidth]{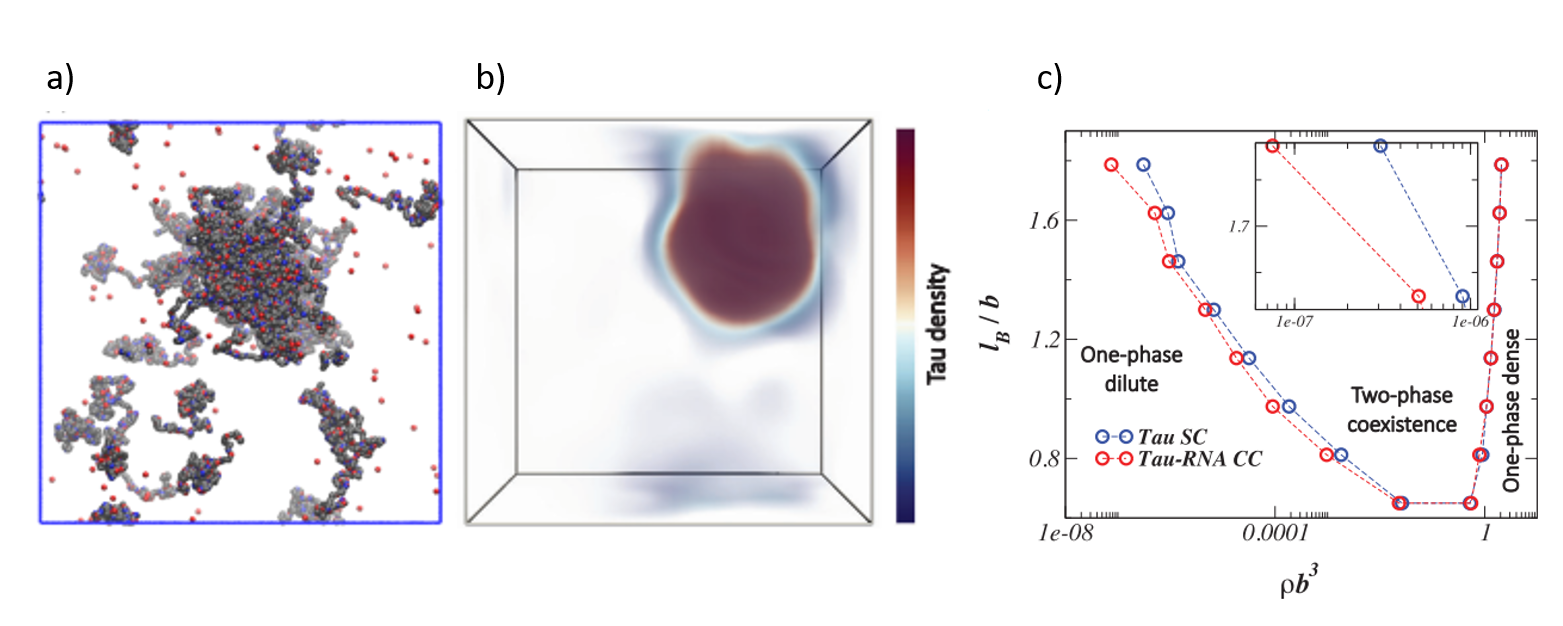}
\caption{Coarse-grained simulations of liquid-liquid phase separated
droplets of Tau proteins \protect\cite{NLLZ_21}.
a) Particle-based CG model using a
Kremer-Grest type representation of chains. b) FTS model, studied
by Complex Langevin simulations. c) Phase diagram obtained from
FTS simulations as a function of Bjerrum length $l_B$ and Tau-density
$\rho$ rescaled with the statistical segment length $b$ (blue curve).
The red curve shows the corresponding phase diagram for mixtures of
Tau and RNA. Adapted from Ref.\ \protect\cite{NLLZ_21} with
permission of XXX}
\label{fig:NLLZ_21}
\end{figure*}

\subsection{Mesoscopic scale: Transition to field theories}
\label{sec:scales_scft}

The next level is the scale of {\em mesoscale organization}, i.e.,
structure formation in inhomogeneous polymer systems.  Emerging
phenomena at this scale are the nucleation of crystallites in
semicrystalline polymers \cite{handbook_polymer_crystallization_13},
phase separation and demixing \cite{handbook_multiphase_polymer_11},
wetting phenomena, or self-assembly \cite{ME_12}.

Apart from the coarse-grained polymer models with hard or soft
interactions discussed in \ref{sec:scales_polymers} and
\ref{sec:scales_blobs}, a new tool for investigating polymer systems on
such scales are {\em field-theoretic approaches} \cite{fredrickson_book,
DF_16, Review_Matsen_20}. The most common starting points for the
derivation of such approaches are density-based model such as
(\ref{eq:potential_density}). By field theoretic manipulations such as
delta functional transformations \cite{Review_Schmid_98,
Review_Schmid_11} or Hubbard-Stratonovich
transformations \cite{fredrickson_book}, one can rewrite the partition
function of this system as an integral over fluctuating real and
imaginary fields.  For example, the delta functional transformation of
the model (\ref{eq:potential_density}) yields the following expression
for the partition function \cite{Review_Schmid_98}:
\begin{equation}
\label{eq:z_fts}
{\cal Z} \propto
\int_{i \infty}  {\cal D} \underline{W} 
\int_\infty {\cal D} \underline{\rho} \:
\exp(-{\cal F}/k_B T)
\end{equation}
with
\begin{eqnarray}
\label{eq:F_fts}
\frac{{\cal F}[\underline{\rho},\underline{W}]}{k_B T} 
&=& \int \ud^3 r \; {f(\vec{r},\underline{\rho})}\Big/{k_B T}
\\ \nonumber &&
  - \sum_\alpha \int \ud^3 \rho_\alpha W_\alpha
  - \sum_j n_j \ln({\cal Q}_j/n_j),
\end{eqnarray}
(in the canonical ensemble) where $\underline{W} = \{W_\alpha\}$
denotes a vector of fluctuating imaginary auxiliary fields
$W_\alpha(\vec{r})$, $j$ sums over different polymer types, and ${\cal
Q}_j[\underline{W}]$ is the single-chain partition functions of
polymers of type $j$ {\em without} non-bonded interactions in the
fluctuating external field $\underline{W}$. 

Taking these expressions as a starting point, one can make several
approximations: First, one can replace the integral (\ref{eq:z_fts}) by
a saddle point approximation, which amounts to approximating the free
energy of the system by the extremum of ${\cal F}$ in Eq.
(\ref{eq:F_fts}). Remarkably, the extremum for $\underline{W}$ is not on
the original imaginary integration domain, but purely real. The
approximation results in the so-called {\em self-consistent field} (SCF)
theory \cite{Review_Schmid_98, Review_Schmid_11, Review_Matsen_02}, one
of the most powerful mean-field approximation for inhomogeneous polymer
systems, which can often predict real interfacial structures in polymers
at an almost quantitative level \cite{Review_Schmid_11}.  Figuratively
speaking, the SCF theory describes polymer systems as assemblies of
independent chains, each in the ensemble-averaged field of the
surrounding chains. The averaging approximation is good if chains
interact with many other chains, which is true for chains of high
molecular weight since they overlap with each other. In three
dimensional melts of linear polymers, the degree of interchain
interactions can be characterized by the so-called invariant
polymerization index $\bar{N}=a^6 \rho_0^2 N$, where $a$ is the average
segment length, and $N$ the numbers of segments in a polymer chain. For
$\bar{N} \to \infty$, the SCF approximation becomes exact.
Experimentally relevant values of $\bar{N}$ are of order $10^2-10^4$.  

A second approximation to (\ref{eq:z_fts}) consists in applying a
partial saddle point approximation only with respect to the auxiliary
fields $W_\alpha$. Thus the functional ${\cal
F}[\underline{\rho},\underline{W}]$ is extremized with respect to
$\underline{W}$, giving self-consistent equations for
$\underline{W}[\underline{\rho}]$, which have again, a real solution for
$\underline{W}$. This procedure turns ${\cal F}$ into a real-valued
density functional ${\cal F}[\underline{\rho}]$.  It serves as starting
point for dynamic mean field theories of polymers which have the
structure of continuum theories, but retain some knowledge of the
macromolecular architecture of polymers.  The simplest Ansatz of this
kind is the purely diffusive equation of motion \cite{KS_87, KS_88,
Fraaije_93, FVMP_97, MF_97, KDH_99, MKD_01, RMB_01,
Review_MS_05,QS_17, MQS_20, SL_20}
\begin{equation}
\label{eq:ddft}
\partial_t \underline{\rho}(t)
 = \nabla_r \int \ud^3 r \underline{\underline{\Lambda}}(\vec{r},\vec{r}')
\nabla_{r'} \underline{\mu}(\vec{r}',t)
\end{equation}
with $\mu_\alpha(\vec{r},t) = \delta {\cal F}/\delta
\rho_\alpha(\vec{r},t)$. Here the mobility matrix function
$\underline{\underline{\Lambda}}(\vec{r},\vec{r}') = \{\Lambda_{\alpha
\beta} (\vec{r},\vec{r}')$ describes the motion of monomers at position
$\vec{r}$ in response to a a local thermodynamic force $(- \nabla_{r'}
\underline{\mu}(\vec{r}',t)$ and may be nonlocal to account for the
effect of chain connectivity. Possible extensions include the coupling
to equations of fluid dynamics in order to account for
hydrodynamics \cite{HK_08,MZSV_98,ZSS_11,HSS_17}, or the introduction of
a time-delayed response functions to account for
memory \cite{WRM_19,RM_20}. 

Going beyond the mean-field approximation, {\em field-theoretic
simulations} (FTS),  aim at sampling the full partition function
(\ref{eq:z_fts}). The field of FTS is relatively new and, so far,
restricted to static simulations. An important problem that needs to
be overcome is the imaginary integration domain of $\underline{W}$ in
the integral (\ref{eq:z_fts}). Since the $W_\alpha$ are imaginary, the
''action'' ${\cal F}$ in (\ref{eq:F_fts}) is a complex, rapidly
oscillating quantity, which leads to a sign problem.  Pioneered by
Ganesan and Fredrickson \cite{GF_01}, one approach to overcoming this
problem is to use the so-called ''Complex Langevin'' simulation
method, which involves solving Langevin equations in the entire
complex plane, in the case of Eq.\ (\ref{eq:z_fts}) for both the
$\underline{W}$ and $\underline{\rho}$ degrees of freedom. 

If the underlying non-bonded potential functional is a quadratic
functional of the densities in $\underline{\rho}$ as, e.g., in Eq.\
(\ref{eq:F_fts}), one can reduce the number of fluctuating fields by a
factor of two by applying a Hubbard Stratonovich transformation
instead of a delta functional transformation, which significantly
reduces the computational costs.  Complex Langevin simulations based on
this approach have been used by Fredrickson and coworkers and other
groups to study, among other, fluctuation effects in diblock copolymer
phase diagrams \cite{GF_01,DF_16}, polymer nanocomposites \cite{LCR_19},
polyelectrolyte complexation \cite{DF_16}, and liquid-liquid phase
separation of intrinsically disordered  proteins \cite{MDDF_19,NLLZ_21}. 
Fig.~\ref{fig:NLLZ_21} shows a FTS simulation droplets formed from
Tau-Proteins, strong polyampholytes which undergo liquid-liquid
phase separation due to self-coacervation.

\begin{figure}[t]
\includegraphics[width=0.4\textwidth]{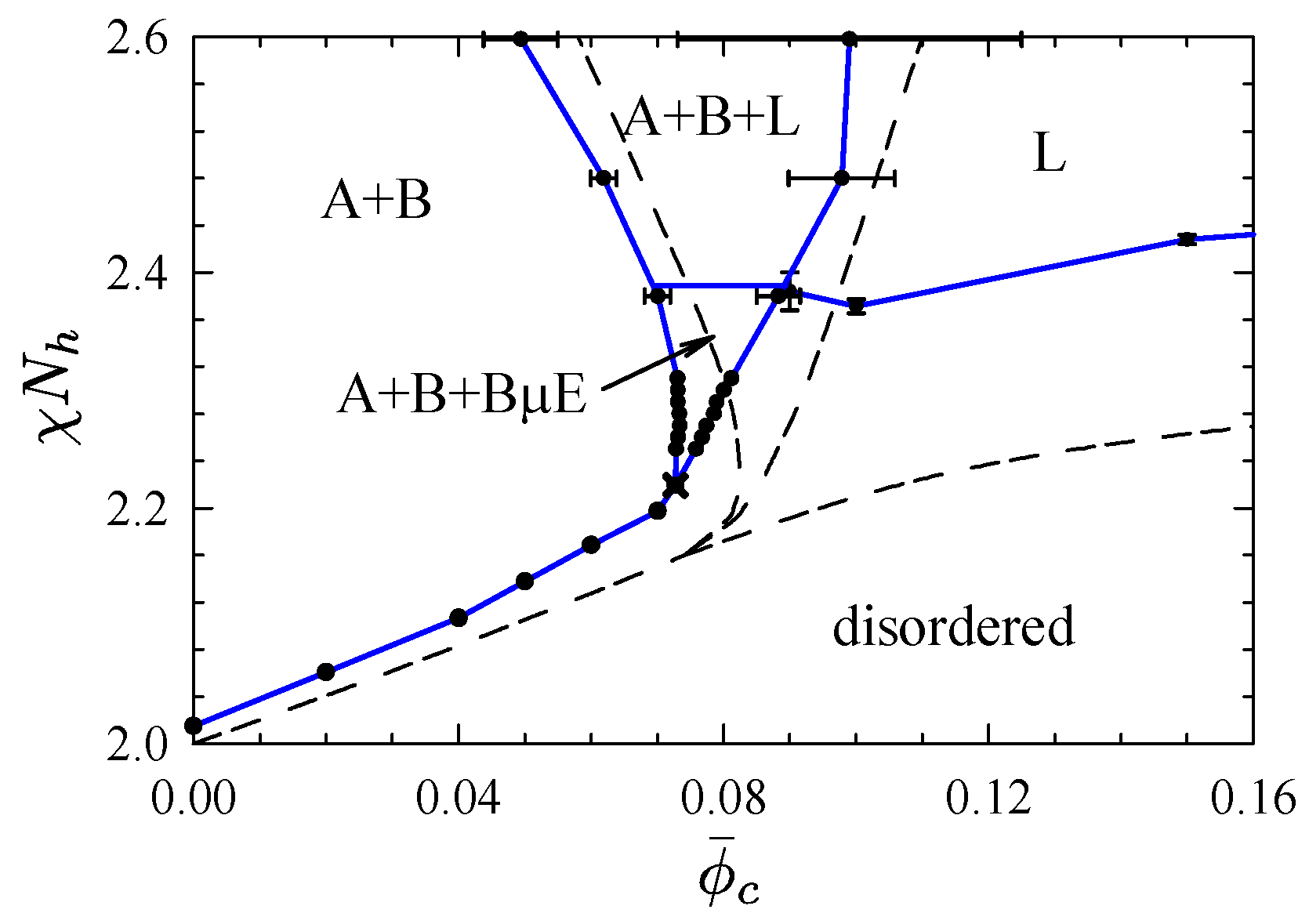}
\caption{
Fluctuation effects on the Lifshitz point in the phase diagram 
of ternary  A/B/AB homopolymer/diblock melts, as revealed by
field-theoretic simulations
in Ref.~\protect\cite{VSM_20, SM_21}. Dashed line shows mean-field 
result, symbols connected by by solid lines the simulation results.
Reprinted from Ref.\ \protect\cite{Review_MB_21}  with
permission of XXX}
\label{fig:Review_MB_21}
\end{figure}

In dense melts of polymers containing only two types of monomers A and
B, a second approach becomes possible \cite{Review_MS_05, DGFS_03,
DS_04, RMB_01, Review_Matsen_20, Review_MB_21}, which has been applied
with considerable success by Matsen and coworkers to study block
copolymer systems \cite{Review_Matsen_20}: In that case, two
fluctuating fields result from the Hubbard-Stratonovich
transformation, an imaginary one which can be associated with density
fluctuations, and a real one which describes the composition
fluctuations. In nearly incompressible melts, the density fluctuations
have little influence on the composition fluctuations, which determine
the phase behavior. Therefore, one may apply a partial saddle point
approximation regarding the density fluctuations only, and obtains a
purely real fluctuating field theory, which can be treated, e.g., by
standard Monte Carlo methods.  Comparisons with Complex Langevin
simulations \cite{DGFS_03,AF_07} have shown that the partial saddle
point approximation is indeed accurate in dense melts.  The advantage
of the approach is that it allows more easily to access highly
incompressible melts at experimentally relevant polymerization
indices \cite{Review_Matsen_20}. 

In many cases, fluctuation corrections mainly shift phase transition
temperatures or change the order of a transition from second order to
weakly first order, but there are situations where they may
fundamentally change the properties of a system. One prominent example
is the ''microemulsion channel'' in balanced mixtures of A,B
homopolymers and A:B diblock copolymers.  Upon increasing $\chi N$,
SCF calculations predict a demixing transition at low copolymer
content, and an ordering transition to a periodic lamellar phase at
high copolymer content. According to the SCF theory, both transitions
meet at a so-called ''Lifshitz critical point'', where the lamellar
thickness of the periodic phase diverges.  In reality, generic
theoretical considerations \cite{Zappala_18} suggest that the Lifshitz
critical point has a lower critical dimension of four, meaning that
fluctuations will invariably destroy it in three (or fewer)
dimensions. The fate of the Lifshitz point in lower dimensions has
long remained unclear, but was recently revealed by field-theoretic
simulations of Vorselaers, Spencer and Matsen \cite{VSM_20, SM_21}: It
splits up into a critical end point and a tricritical point (see Fig.\
\ref{fig:Review_MB_21}).  This example also demonstrates how
simulations of polymer systems can give insights onto fundamental
questions in statistical mechanics.

\begin{figure*}[t]
\includegraphics[width=0.9\textwidth]{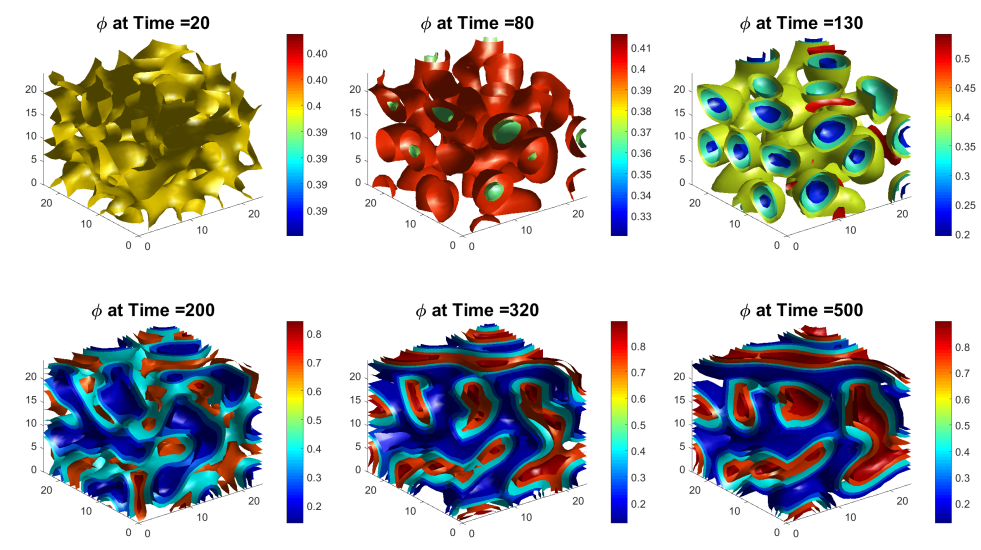}
\caption{Spinodal phase separation in a continuous viscoelastic model.
The numerical simulation is based on a Lagrange-Galerkin finite element method. 
Adapted from Reference \protect\cite{BL_22b} with permission of XXX.}
\label{fig:viscoelastic} \end{figure*}

\subsection{Macroscopic scale: The engineering scale}
\label{sec:scales_macroscopic}

Finally, at the {\em macroscopic} level, the focus lies on properties
of polymeric materials that are of direct interest for engineers:
Mechanical stability, microstructure, stress distribution,
viscoelasticity, constitutive relations, ageing phenomena. Emerging
phenomena that are studied on such scales are, for instance, polymeric
flow patterns in complex geometries \cite{han_book}, but also
inherently inhomogeneous processes such as viscoelastic phase
separation \cite{Tanaka_00}, foaming \cite{GP_15,obi_book}, or crack
formation \cite{heinrich_book}.  On macroscopic scales, materials are
described by a set of characteristic continuous fields and
corresponding transport equations. They are typically constructed
somewhat heuristically based on general symmetry considerations and
conservation laws, following the spirit of the famous
Hohenberg-Halperin classification of dynamic critical phenomena
\cite{HH_77}. For example, so-called ''model A'' dynamics is used to
describe relaxation processes where conservation laws are not
important, ''model B'' dynamics is used to describe diffusive
processes where only the local conservation of one ''order parameter''
(e.g., the polymer volume fraction) matters (an example is Eq.\
(\ref{eq:ddft})), ''model C'' dynamics describes processes where other
conserved mass densities also become important, and ''model H''
additionally accounts for local momentum conservation and convection
in order to describe stress, flow and hydrodynamic phenomena.  

In polymeric materials, these descriptions are often combined with
simplified microscopic models for viscoelasticity, which account both
for flow and internal relaxation processes.  An important tools in the
construction of such models is the so-called {\em convected
derivative}, a concept originally introduced by Oldroyd
\cite{Review_Beris_21}: It
describes convection with respect to a ''material frame'' of
comoving material particles and thus in some sense generalizes the
total derivatives in fluid dynamics to tensorial quantities: Consider
a fluid flow characterized by a deformation rate field $G_{ij} =
\partial_j u_i$ and carrying a scalar field $\Phi(\vec{r},t)$ and a
tensor field $Q_{ij}(\vec{r},t)$.  Let $\Phi^{(L)}(\vec{r},t)$ and
$Q^{(L)}_{ij}(\vec{r},t)$ be the corresponding scalar and tensorial
densities in the comoving frame with
$\Phi(\vec{r},0)=\Phi^{(L)}_{ij}(\vec{r},0)$,
$Q_{ij}(\vec{r},0)=Q^{(L)}_{ij}(\vec{r},0)$. The total derivative
is defined such that it describes the evolution of the scalar
field $\Phi(\vec{r},t) = \Phi^{(L)}(\vec{r} - \vec{u} t,t)$ in
the comoving frame, i.e.,
\begin{equation}
\frac{\ud}{\ud t} \Phi := 
\partial_t \Phi^{(L)}
= \partial_t \Phi + \vec{u}\nabla \Phi,
\end{equation}
When generalizing this concept to the tensorial field $\mathbf{Q}$,
one must take into account the possible deformations of the
coordinates of $\mathbf{Q}$ in the comoving frame, which can be
described by a transformation matrix defined by the deformation rate
$\mathbf{G}$.  For contravariant tensors $\mathbf{Q}$, the relation
between $\mathbf{Q}$ and $\mathbf{Q}^{(L)}$ in the limit $t \to 0$ is
given by $\mathbf{Q}(\vec{r},t) = (\mathbf{1} + \mathbf{G}^T t)
\mathbf{Q}^{(L)}(\vec{r}-\vec{u} t,t) (\mathbf{1}+ \mathbf{G}t)$.
This motivates the definition of the upper convected derivative
\begin{equation}
\overset{\smalltriangledown}{\mathbf{Q}} :=
\partial_t \mathbf{Q}^{(L)}
= \partial_t \mathbf{Q} + \vec{u}\nabla \mathbf{Q}
- \mathbf{G}^T \mathbf{Q} - \mathbf{Q} \mathbf{G}
\end{equation}
The corresponding consideration for covariant tensors yields
the lower convected derivative
\begin{equation}
\overset{\smalltriangleup}{\mathbf{Q}} :=
\partial_t \mathbf{Q} + \vec{u}\nabla \mathbf{Q}
+ \mathbf{G}^T \mathbf{Q} + \mathbf{Q} \mathbf{G}
\end{equation}
The concept of convective derivatives provides a framework for
deriving constitutive relations in viscoelastic materials in a
geometrically consistent manner. For example, the so-called upper 
convected Maxwell modell
\begin{equation}
\mathbf{\sigma} + \lambda
\overset{\smalltriangledown}{\mathbf{\sigma}} = 2 \eta \mathbf{D}
\end{equation}
with $\mathbf{D}=\frac{1}{2}(\mathbf{G} + \mathbf{G}^T)$ 
describes a material with a linear steady-state stress-strain 
relation $\mathbf{\sigma}_{\mbox{\tiny} steady} = 2 \eta \mathbf{D}$
where the stress tensor, $\mathbf{\sigma}$,
relaxes in a simple exponential manner towards its steady-state value 
with relaxation time $\lambda$.

More sophisticated viscoelastic models are typically based on one of
two approaches \cite{Review_BG_16}: Either use phenomenological
considerations to construct more complicated expressions for the
relaxation of the stress tensor and/or its steady-state value, or
select a simplified molecular model of the polymers and use kinetic
theory to derive approximate expressions for the stress tensor. One
popular starting point of the second kind is to consider a Newtonian
fluid filled with noninteracting elastic dumbbells, i.e., two beads by
''Finitely Extensible Nonlinear Elastic'' (FENE) springs.  with a
spring constant $k(R)$ that diverges if the distance $R$ of the beads
exceeds a limiting value.  A Fokker-Planck equation for the
conformation of the dumbbells is then coupled to the Navier-Stokes
equations via two convective contributions to the Fokker-Planck
equation (one for the center of mass and one for the relative distance
of beads) and an extra stress term in the Navier Stokes equations.
Many macroscopic models for polymer fluids can be seen as being
approximations to this FENE model.  The most prominent one is the
Oldroyd-B model, one of the first models for polymer fluids
\cite{Review_Beris_21}, which replaces the FENE spring by a regular
linear Hookean spring \cite{Masmoudi_11}. This simplifies the
mathematical analysis, however, it leads to unphysical singularities
under certain flow conditions where the dumbbells stretch to infinity.
Another approximation is the Peterlin model (FENE-P), where the
nonlinear spring constant $k(R)$ is replaced by an averaged value
$k(\langle R \rangle)$ \cite{Peterlin_66a,Peterlin_66b}. 

In order to avoid mistakes when constructing such models, considerable
care has to be taken to ensure that they are thermodynamical
consistent \cite{degroot_book,SC_21}. Several mathematical frameworks
have been developed which help to enforce consistency, the most
rigorous being the GENERIC framework that makes a strict distinction
between anti-symmetric reversible and symmetric irreversible
(dissipative) contributions to the dynamical equations
\cite{oettinger_book}. It should be noted that not all published
macroscopic models are thermodynamically consistent.  Schieber and
Cordoba have recently developed a simplified set of requirements that
allows one to perform basic consistency tests without having to apply
the full GENERIC machinery \cite{SC_21}.  Another, even more difficult
question, is to prove that the models actually have solutions for
arbitrary initial conditions.  Global existence results for weak
solutions of the Peterlin model have recently been obtained by
Masmoudi \cite{Masmoudi_11} and, regarding a class of generalized
Peterlin models, by Lukacova-Medvidova et al \cite{LMNR_17, GLMS_18}. 

The study of inhomogeneous polymer solutions is particular challenging
due to the vastly different mobilities of polymer and solvent
molecules.  Quite generally, large dynamical asymmetries between
components of a demixing system often result in unconventional
network-like pattern formation and novel dynamic scaling exponents
\cite{TT_21} compared to standard model B or model H demixing, because
the domains of the slow phase tend to behave like viscoelastic
objects. This phenomenon was first discovered by Tanaka in 1993
\cite{Tanaka_93} who termed it ''viscoelastic phase separation'', and
is still subject of active research \cite{Tanaka_96, TO_96, Tanaka_97,
Tanaka_00, ZZE_06,SBHE_21,BDEH_21}. Theoretical models typically build
on the two-fluid model proposed by Doi and Onuki \cite{DO_92} and
Milner \cite{Milner_93}, which include a coupling between elastic
stress and concentration. Based on this idea, Zhou et al proposed a
number of phenomenological models for viscoelastic phase separation,
paying particular attention to thermodynamic consistency
\cite{ZZE_06}. Spiller et al \cite{SBHE_21} have recently taken the
kinetic approach and derived a two-fluid model for solutions of
Hookian dumbbells which is consistent with the GENERIC formalism.
Brunk et al have analyzed a number of models for viscoelastic phase
separation from a mathematical point of view and proved the existence
of weak solutions \cite{BDEH_21,BL_22a,BL_22b}. An example of a
numerical simulation of one of their models is shown in Fig.\
\ref{fig:viscoelastic}.


\section{Scale-bridging strategies}
\label{sec:multiscale}

In the previous section, we have discussed the hierarchy of 
models that have been designed and used to study polymeric systems on
different scales. In many cases, however, using a single model is not
sufficient to fully characterize a material of interest. Thus
multiscale modeling techniques must be applied, which combine
different scales in one simulation, or at least establish quantitative
connections between different scales. The key to multiscale
modeling is coarse-graining, i.e., the art of designing high-level
models with few degrees of freedom (''coarse-grained (CG) models'')
that capture the essential features of an underlying ''fine-grained
(FG)'' system.

Classical coarse-graining strategies traditionally follow one of two
philosophies \cite{Review_Noid_13}: ''Top-down'' CG models are
designed based on physical intuition without direct input from FG
simulations.  Examples are generic top-down models such as the Bond
Fluctuation Model \cite{CK_88} and the Kremer-Grest model \cite{GK_86}
discussed in Sec.\ \ref{sec:scales_polymers}, which are used to study
generic properties of polymer systems, but also chemically specific
top-down models such as the MARTINI model \cite{MRYT_07}, which use
experimental information such as solubility parameters to match
interaction parameters. In contrast, ''bottom-up'' CG models are
constructed from FG simulations in a systematic manner such that they
capture certain structural or thermodynamic properties of interest.
This is a popular strategy to derive classical atomistic force fields
from electronic structure calculations, and it is also used to
construct higher-level models. In addition to bottom-up and top-down
approaches, hybrid approaches are becoming increasingly popular that
integrate information from different sources -- FG simulations as well
as experiments \cite{BCCV_16, RKB_16, FBCB_20, MMKU_21}, and
data-driven methods that apply machine-learning methods \cite{MMKU_21,
Review_NTMC_20, Review_Webb_21}.

Numerous coarse-graining and scale-bridging strategies have been
proposed over the past decades (see Refs.~\cite{Review_Peter_10,
Review_Noid_13, Review_Liu_13, Review_vanderVegt_13, Review_Voth_13,
Review_BG_16, Review_Rudzinski_19, Review_Memory_21,Review_Webb_21,
NTL_22} for review articles), and giving a comprehensive overview is
beyond the scope of the present perspective article. Instead we
will give a very personal view on different aspects of the
coarse-graining problem with a focus on bottom-up coarse-graining, on
lessons learned from the past and challenges for the future. 

Formally, defining the coarse-graining task seems quite obvious: Given
a microscopic dynamical system with $N$ degrees of freedom and
corresponding equations of motion, define a reduced set of $n$
representative collective variables and derive their dynamical
equations from those of the microscopic system. This idea is old and
projection operator techniques to derive coarse-grained equations have
been proposed already in the 60s by Zwanzig and Mori \cite{Mori_65a,
Mori_65b,Zwanzig_61}. They were used, among other, to derive equations
of fluctuating hydrodynamics for simple and complex fluids
\cite{forster_book,boon_yip_book}.  In recent years, the Mori-Zwanzig
formalism has attracted increasing interest in the coarse-graining
community, mostly thanks to the work of Espa{\~{n}}ol and coworkers
who promoted it as a practical tool to construct, e.g., dynamic
density functional theories \cite{EL_09} or particle-based DPD models
\cite{HEVD_10}. In principle, projection operators allow one to derive
{\em exact} dynamical equations for the chosen coarse-grained
variables. However, these are complex integro-differential equations
that cannot be reduced to practically useful model equations, e.g.
stochastic equations, without substantial further approximations. Even
more seriously, Glatzel and Schilling have recently argued that the
dynamic equations for the coarse-grained variables $A_i(t)$ cannot
necessarily be related to a potential of mean force \cite{GS_21,
Review_Schilling_22} $U[A_i]$. Their claim is consistent with a
discussion by Zwanzig in Ref.\ \cite{zwanzig_book}, who pointed out that the
memory kernel in the linear Mori-Zwanzig equations absorbs some of the
nonlinearities of a nonlinear conservative potential in the FG
equations. Unfortunately, it implies that the resulting coarse-grained
models are not necessarily compatible with the GENERIC framework
\cite{oettinger_book} and its clear distinction between external
driving forces, conservative interactions, and dissipative forces. As
discussed earlier, the GENERIC structure helps to enforce
thermodynamic consistency and ensure, by construction, that violations
of the second law of thermodynamics are not possible in a
coarse-grained model. Giving up this structure thus represents a
serious drawback. Luckily, recent work by Vroyland and Monmarch\'e
suggests a possible way out of this dilemma. Using the Mori-Zwanzig
formalism and considering a single CG particle, they showed that it is
possible to derive a GLE that complies with the GENERIC structure, if
one allows for position dependent memory kernels\cite{VM_22}. 

One may be tempted to set aside these problems and design CG models
that primarily target static equilibrium properties. One can then use
the partition function of the microscopic system as the starting point
and integrate out the $N$ microscopic degrees of freedom while
constraining the $n$ CG variables, which directly gives the potential
of mean force or ''free energy landscape'' $U[A_i]$. In general,
however, simple analytic expressions for $U[A_i]$ are not available,
such that a simulation of the exact CG model is as expensive, from a
computational point of few, than the simulation of the FG model. 
Thus further approximations must again be made such as, e.g.,
rewriting $U[A_i]$ as a sum of effective pair or low order multibody
potentials.

Finally, already the identification of meaningful coarse-grained
variables represents a challenge in itself -- in particular if the
coarse-grained model is expected to capture several very different
aspects of the underlying FG model. This leads to the
well-known problem of {\em representability}: A
CG model that reproduces the structure of the FG model does not
necessarily have the correct thermal properties and vice versa. 
Moreover, a CG model that was constructed for one state point (e.g.,
one density), not necessarily captures the properties of the FG model
at another state point (another density). This so-called {\em
transferability} issue will obviously cause problems when using CG
models for studying strongly inhomogeneous systems.

In sum, coarse-graining is bound to be a somewhat ''dirty'' business.
The reason is that, unfortunately, it is not possible to simplify a
complex problem just by rewriting it in terms of fewer variables.
Coarse-graining is effectively an optimisation problem which requires
many compromises and a high level of physical and chemical intuition.
The coarse-graining philosophy rests on the assumption that the
large-scale structure of materials can be understood without explicit
knowledge of microscopic details. In the case of polymers, one hopes
that this assumption is justified due to their repetitive molecular
structure, the high level of conformational disorder, and the dominant
role of entropy.

We will now discuss selected aspects of coarse-graining in polymeric
systems, or more generally, soft matter systems.

\subsection{Static coarse-graining}
\label{sec:multiscale_static}

\subsubsection{Structure-based coarse-graining}
\label{sec:multiscale_static_structure-based}

Structure-based coarse-graining techniques are typically used to
design particle-based CG models with the goal to reproduce structural
properties of the FG system such as spatial correlation functions.
The CG variables are the positions $\mathbf{R}_i$ of CG particles, and
the optimization task consists in finding the best approximation for
the free energy landscape $U[\mathbf{R}_i]$ or the configuration
dependent force field $\mathbf{F}_j[\mathbf{R}_i]$ in the phase space
of the CG variables. Regarding equilibrium static coarse-graining, the
field is already quite advanced. The CG bonded interactions can be
calculated in a straightforward manner by sampling, e.g., bond length
and bond angle distributions in small reference simulations.  To
determine non-bonded CG interactions, researchers can use the
open-source package VOTCA \cite{RJLK_09} (www.votca.org) and select
between a range of established methods \cite{Review_Noid_13,
Review_vanderVegt_13} such as inverse Monte Carlo (IMC) \cite{LL_95},
iterative Boltzmann inversion (IBI) \cite{Soper_95,RPM_03,GMJV_12,
ONKJ_16, Hanke_17}, force matching (FM)
\cite{EA_94,IV_05a,IV_05b,NCAK_08,NLWC_08}, or relative entropy (RE)
minimization between the CG and the FG distribution \cite{Shell_08}.
Alternatively, they can employ the framework of the generalized
Yvon-Born-Green (g-YBG) \cite{MN_09, MN_10, WBKR_19}, or use
artificial neural networks \cite{WOWP_19,Review_NTMC_20,BN_21}. Noid
et al have pointed out that the quality of non-bonded CG force fields
can be greatly improved if one distinguishes between CG monomers that
have different local connectivities within a molecule \cite{MN_09b,
DN_16}, e.g., between middle and end segments. 

\begin{figure}[t]
\includegraphics[width=0.5\textwidth]{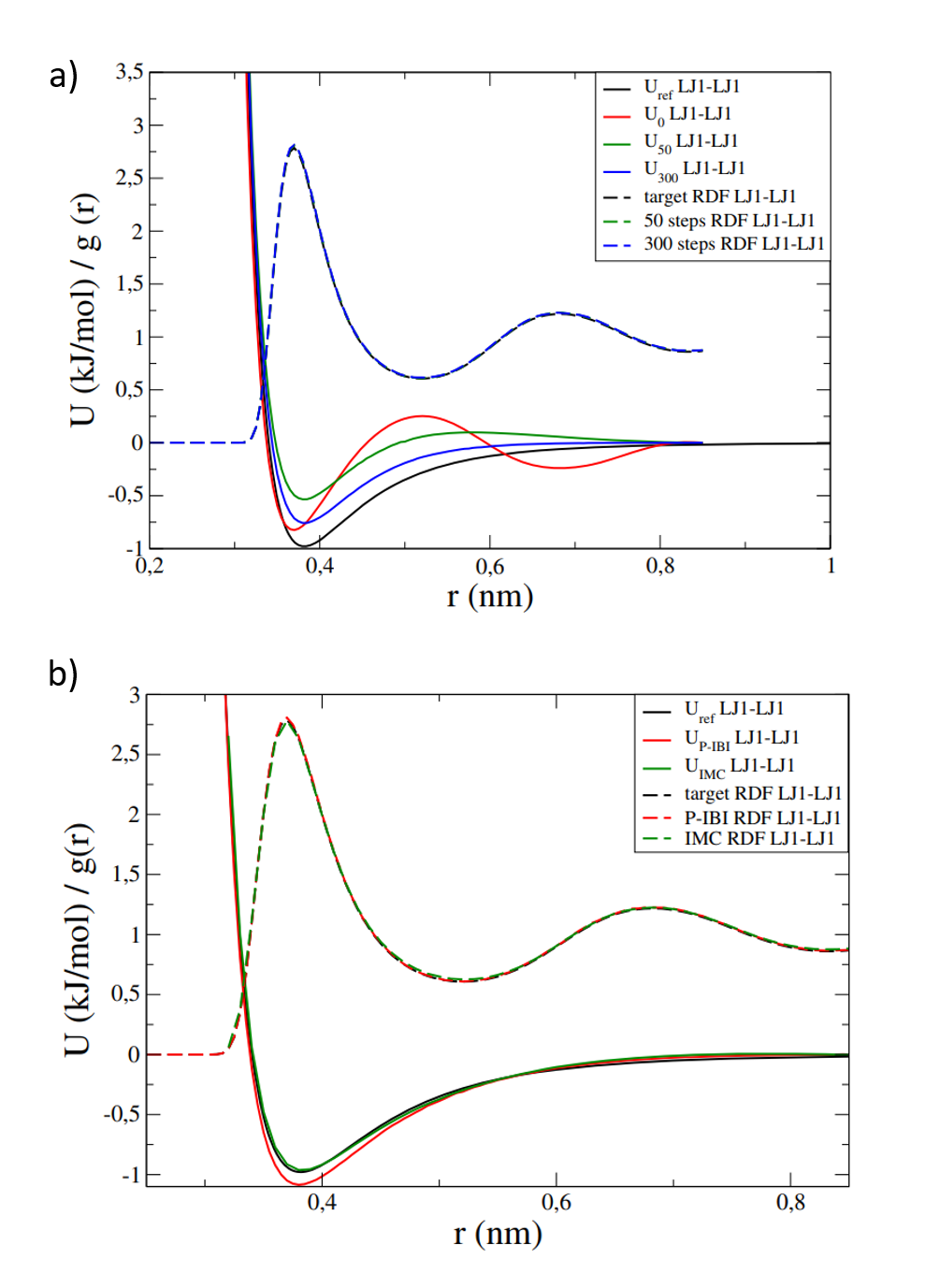}

\caption{Uncertainties in the reconstruction of pair potentials from
pair correlation function. In this example, the target RDF is taken
from simulations of a binary Lennard-Jones mixture, hence the true
potential (black solid line) is well-known.  Solid lines show
potentials as indicated, dashed lines with the same color the RDFs 
obtained with the same potential.
a) IBI results after 50 and 300 iterations (green and blue)
(Red line shows the logarithm of the RDF for comparison.)  
b) Final results obtained with an IBI variant (red) and with IMC
(green). The true potential is best reconstructed with
the IMC method. All potentials, however, yield RDFs that
are almost indistinguishable from the target RDF.
Reprinted from Ref.\ \protect\cite{RHV_16} with permission of XXX.
}
\label{fig:RHV_16}
\end{figure}

One paradigmatic problem in structural coarse-graining is to construct
pair potentials from radial distribution functions (RDFs) of particles
as determined, e.g., from FG simulations. This is known as
the inverse Henderson problem. As proved in 1974 by Henderson for
finite systems with fixed particle number \cite{Henderson_74} and very
recently by Frommer et al \cite{FHS_19, FH_22} for the thermodynamic
limit, the problem has a unique solution for a rich class of
interaction potentials which includes, among other the so-called
Lennard-Jones type potentials \cite{ruelle_book}. Nevertheless, the
problem is ill-posed in the sense that a small noise in the RDF can
lead to large changes in the potentials. In other words, quite
dissimilar potentials can produce almost identical RDFs \cite{VPV_10,
RHV_16} (see also Fig. \ref{fig:RHV_16}).  This opens possibilities to
optimize pair potentials not only with respect to structural
properties, but also to other properties as well. Building on this
idea, Hanke and coworkers have recently developed novel integral
equation-based methods that allow to solve the inverse Henderson
problem with additional constraints \cite{DHI_20,BHV_21}, such that
the resulting CG model reproduces both the structural correlations and
the thermodynamic properties of the microscopic system \cite{BHV_21}. 

An interesting alternative way of dealing with the representability
problem has been proposed by Lebold and Noid \cite{LN_19a,LN_19b}.
Rather than trying to find one CG model that captures both the
energetics and the structure of the FG model, they suggest to
explicitly keep track of energetic and entropic contributions to the
potential of mean force \cite{LN_19a,LN_19b} in the CG simulation. Thus
the effective potential is split up as \begin{equation}
U[\mathbf{R}_i] = U_W[\mathbf{R}_i] - T \: S_W[\mathbf{R}_i],
\end{equation} where $U_W$ is constructed such that it gives, on
average, the energy of the {\em fine-grained} system with collective
variables constrained to $\mathbf{R}_i$. The potential $U$ is obtained
by standard structural coarse-graining methods, the potential $U_W$
ist determined using a minimization method similar to least-square
fitting.  When analyzing CG simulation trajectories, the potential
$U_W$ can then be used to calculate observables that depend on energy.
As a side effect, this approach also allows one to estimate the
expected change in the potential of mean force at a different
temperature with remarkable accuracy \cite{LN_19b}. Among other, this
could be used to overcome sampling problems in the microscopic
reference system, e.g., close to a glass transition \cite{SN_21}.

Going beyond pure pair potentials, higher-order multibody potentials
\cite{MM_09,LLV_10,DA_12,SA_18,GZV_18} or density-dependent
potentials \cite{AR_08,MBIL_16,SS_16, SS_18, DJV_17, DN_17, MKN_17, 
DN_19, RSSV_19, BSAN_21} offer additional flexibility which can be exploited to
develop CG models with improved transferability
properties \cite{JC_17}. In particular, density-dependent potentials
provide a comparatively straightforward way of accounting for the
local environment of interacting CG particles that may undergo
liquid-vapor phase separation \cite{JV_18}, and they are quite popular
in empirical models with soft potentials such as (many-body)
DPD \cite{PF_01} or models with density-based interactions
\cite{HM_10}.  Such empirical soft potential models are often set up
as a sum of two contributions: Local density-dependent repulsive
interactions between particles that account for the effect of excluded
volume interactions, and density-independent attractive interactions
that account for cohesion.  The functional form of the two terms is
usually postulated, but they can also be derived in a bottom-up
fashion from FG simulations, e.g., by a combination of force matching
and relative entropy minimization \cite{DN_19,BSAN_21}.  Since one has
some freedom how to distribute the forces between different
contributions, the results are not unique \cite{DN_19}, they depend on
the coarse-graining procedure. This gives freedom which can be
exploited to further optimize the potentials with respect to
representability and transferability.  

We should note that density-dependent potentials also have interesting
applications beyond liquid-vapor systems, e.g., in
ultra-coarse-grained descriptions of compressible fluids or in
coarse-grained descriptions of responsive materials where the shape of
CG particles depends on their local environment \cite{BD_21}. 

In other situations where local orientations of molecules or monomers
are important, it might be desirable to include multibody
potentials \cite{Review_CWOL_14} that also depend on the local
conformation, such as three-body Stillinger-Weber type potentials that
depend on the local angles between the interaction
sites \cite{SW_85,SW_86, MM_09}.  Scherer, Andrienko, and coworkers have
developed systematic ways to derive such potential, either by force
matching \cite{SA_18} or using kernel-based machine learning with
covariant meshing in order to account for inherent symmetries
\cite{SSAB_20}.  So far, this has only been applied for small
molecules, but it also offers interesting perspectives for simulations
of, e.g., semicrystalline polymers.

An alternative way of parameterizing strongly configuration dependent
effective interaction potentials has recently been developed by Bereau
and Rudzinsky \cite{BR_18, RB_20}. Their idea is to use different CG
force fields for different regions in (local) configurational space,
and interpolate ('hop') between them depending on the current state of
the system. This multisurface concept, borrowed from models for
electronic transitions, allows for a much better local optimization of
force fields without having to resort to complicated force field
parametrizations. As a nice side effect, potential barriers can also
be represented much more accurately, which greatly improves the
dynamic properties of the system. A related ''multiconfigurational''
concept designed to capture conformational chain transitions and their
effect on potentials of mean force has been proposed by Sharp et
al \cite{SVWD_19}.

\subsubsection{Thermodynamics-based coarse-graining}
\label{sec:multiscale_statics_TD-based}

The coarse-graining approaches discussed in the previous section
yield CG models that capture structural properties of the FG
reference system such as pair correlation functions or statistical
averages of mechanical force fields. From a multiscale point of view,
it might often be more interesting to capture thermodynamic properties
such as the equation of state (the density), the compressibility or
more generally, high-wavelength structure factors, solubility
parameters, interfacial tensions and surface tensions.
Thermodynamics-based CG strategies are popular in top-down
coarse-graining, as they use thermodynamic quantities as input which
are more easily accessible in experiments. Typically, a certain
functional form of potentials is assumed, and the parameters are
matched such that the CG model reproduces the desired thermodynamic
properties.


Thermodynamics-based coarse-graining is also the most natural approach
when designing field-based continuum model s or extremely CG models
with soft potentials such as DPD or density-based potentials. As we
have discussed above, density-based models and field-based continuum
models are closely related to each other. There also exists a direct
connection between DPD and continuum mechanics: For simple fluids,
Espa{\~{n}}ol and Revenga have introduced a variant of DPD
\cite{ER_03}, termed ''smoothed dissipative particle dynamics''
(sDPD), which is entirely constructed from thermodynamic properties
and can be seen as a Lagrangian solver for the fluctuating
Navier-Stokes equations. 

When constricting ultra CG models, one must again distinguish between
bonded and non-bonded potentials.  Bonded potentials can be determined
in a structure-based manner as described in Sec.
\ref{sec:multiscale_static_structure-based}.  On large scales, when
studying polymers of large molecular weight, it is often sufficient to
use simple chain models such as the discrete or continuous Gaussian
chain model \cite{doi_edwards_book}, which requires matching only one
parameter (the Kuhn length) \cite{flory_book} to the average
conformational properties of the chains in the reference system.
Determining non-bonded interactions is more difficult, as standard
density-based potentials or interaction terms in field-based models
(Eqs.\ (\ref{eq:potential_density}) and (\ref{eq:F_fts})) are
typically framed in a thermodynamic language in terms of
compressibilities, Flory Huggins $\chi$ parameters (Eq.\
(\ref{eq:f_rho})) etc. 

Specifically, mapping $\chi$ parameters still represents an
outstanding challenge.  Field-theoretic models typically assume that
it describes the effective non-bonded interactions between CG polymer
segments and is independent of local composition, chain length and
chain architecture.  This picture is clearly greatly simplified and it
has long been unclear whether the concept of a purely monomer-based
$\chi$ parameters is at all reasonable. Luckily, recent work by Morse,
Matsen and coworkers \cite{Review_Matsen_20} on diblock copolymer
melts suggests that this is probably the case, at least for dense
polymer melts, due to a  universality in the phase behavior of
polymers with large molecular weight.  They proposed to determine the
$\chi$ parameter in diblock copolymer melts by fitting the collective
structure factor in the disordered state of symmetric diblock
copolymer (BCP) melts to accurate theoretical predictions of a
renormalized theory that accounts for the effect of fluctuations and
finite chain lengths \cite{GQMM_14}. Using this top-down mapping
method, they were able to quantitatively reproduce the location of the
order-disorder transition (ODT) in BCP melts of a number of
particle-based models \cite{GQMM_14, GMBM_14} and also experimental
systems \cite{BM_16} after accounting for the effect of
polydispersity. Building on this insight, Willis et al \cite{WBM_20}
proposed as alternative approach to directly use the ODT for mapping
$\chi$ after correcting for effects of polydispersity and
compositional asymmetry.  Reanalyzing published experimental data,
they mapped $\chi(T)$ onto the functional form

\begin{equation}
\chi(T) = \frac{A}{T} + B,
\end{equation}
where $A$ subsumes energetic and $B$ entropic contributions to the
effective segment interactions, and extracted values of $A$ and $B$
for 19 different chemical pairs.

The $\chi$-calibration scheme of Morse and coworkers relies on the
existence of accurate theoretical predictions for the structure and
phase behavior of diblock copolymer melts. When looking at more
complex systems, such predictions are usually not available, and less
accurate mapping procedures must be adopted. A number of heuristic
schemes for matching Flory Huggins type interaction parameters have
been proposed by De Nicola et al \cite{NZKR_11, NKM_13}, that either
rely on matching energies with CG off-lattice models  \cite{NZKR_11} or
adjusting conformational properties of homopolymers in
solution \cite{NKM_13}. Ledum et al have developed a machine-learning
protocol for optimizing such parameters with respect to arbitrary
target quantities, e.g., density profiles \cite{LBC_20}. 

Sherck et al \cite{SSNY_21} and Weyman et al \cite{WMO_21} have
recently developed systematic bottom-up coarse-graining strategies for
deriving field-based models with non-bonded monomer interactions that
are not restricted to the Flory Huggins form (\ref{eq:f_rho}).
Their idea is to proceed in two steps. In the first step, a CG
particle-based with soft pair potentials of given functional form is
determined from reference simulations of a microscopic model, e.g., by
relative entropy minimization \cite{SSNY_21} or by matching the RDF.
The second step is a Hubbard stratonovich transformation that turns
the particle model into an auxiliary field model, which can then be
studied by field-theoretical simulations. The second step involves an
inversion of the pair potential which is not possible for hard core
potentials, therefore the first step is essential and cannot be
omitted. This still remains true if one replaces the
Hubbard-Stratonovich transformation by a delta functional
transformation in order to obtain a field theory of the type
(\ref{eq:z_fts}, \ref{eq:F_fts}). The underlying density-based
potential does not have to be an integral over a local free energy
density $f(\vec{r},\underline{\rho})$, it could also describe nonlocal
interactions as, e.g., in 
\begin{displaymath}
U_{\mbox{\footnotesize nb}}[\underline{\rho}] 
=  \frac{1}{2}\sum_{\alpha \beta} \iint \ud^3 r \: \ud^3 r' \: 
    \rho_\alpha(\vec{r}) \: \rho_\beta(\vec{r}') \:
     V_{\alpha \beta}(\vec{r} -\vec{r}'),
\end{displaymath}

however, $V_{\alpha \beta}(\vec{r})$ then still needs to be
integrable.  We should note that, strictly speaking, the bottom-up
approaches of Sherck et al and Weyman et al use ideas taken
from structural coarse-graining.  Nevertheless, the resulting CG
models do not capture local structural properties such as packing
effects, hence they are closer to thermodynamically CG models than to
structurally CG models.

Compared to structural coarse-graining, one disadvantage of
thermodynamics-based coarse-graining is that one loses the direct
connection between mechanical forces in the CG and the FG system.
Since forces drive the dynamics, it becomes more difficult to restore
the correct dynamical properties without further adjustments.  Indeed,
recent studies on ionic liquids \cite{RKWP_21} have suggested
that structure-based CG models tend to have a more consistent
dynamical behavior than thermodynamically CG models, e.g., regarding
the relative mobility of anions and cations. We will specifically
discuss issues of dynamic coarse-graining in the next section.
Thermodynamics-based coarse-graining also becomes questionable in
systems far from equilibrium, e.g., in active fluids. One should note,
however, that most structure-based coarse-graining techniques are also
no longer applicable for such systems, as most of them -- with the
exception of force matching -- assume local thermodynamic equilibrium. 

\subsection{Dynamic coarse-graining}
\label{sec:multiscale_dynamics}

The most common approach to studying dynamical properties in CG
simulations is to use the free energy landscape obtained from a static
equilibrium coarse-graining procedure as an effective interaction
potential in molecular dynamics (MD) simulations. This approach can be
quite successful if one accounts for a few side effects of structural
coarse-graining: First, as known from the Mori-Zwanzig formalism
\cite{zwanzig_book} integrating out degrees of freedom invariably
introduces friction and stochastic noise in the CG dynamical
equations. In a standard MD simulation, these friction terms are
disregarded, which accelerates the dynamics. Second, CG free energy
landscapes are typically smoother than atomistic ones, which further
reduces the direct friction between CG particles. As a result,
coarse-graining reduces the separation between originally highly
disparate time scales such as, e.g., the inertial and the diffusive
time scales (the telescope effect \cite{PL_06}, and accelerates slower
dynamical processes. 

From a point of view of time bridging, both the speedup and the
telescoping are beneficial, as they allow one to access later time
scales in CG simulations \cite{KBVV_21} and study processes on
different time scales simultaneously in one simulation.  One of the
earliest \cite{TKBB_98} and still widely and successfully used
approaches to dynamic coarse-graining has been to simply take
advantage of this effect, determine the speedup factor of the process
of interest, and use this to map the CG dynamics on real dynamics
\cite{HAVK_06, HK_09, FHKV_09, FKHV_11}, taking into account that the
speedup factor might be different for different processes and/or
components \cite{FKHV_11}. 

However, care must be taken that the speedup does not change the order
of ''faster'' and ''slower'' processes and which might change
dynamical pathways, particularly in dynamically asymmetric systems.
This defines the problem of {\em dynamic consistency}
\cite{Review_Rudzinski_19}. As already discussed in the previous
section, one key to reducing this problem is accurate structural
coarse-graining, as it helps to recover consistent dynamics even in
standard MD simulations, i.e., consistent barrier crossing dynamics
and consistent relative speedup \cite{BR_18,RKWP_21}. In fact, using
kinetic information as additional input for the parametrization of
coarse-grained force fields may improve their quality \cite{RB_16},
because it gives more weight to transitional conformations, which are
typically not well sampled in standard coarse-graining approaches.  In
addition, one can manually reintroduce terms in the dynamical
equations that mimic the effect of the interactions between the CG
variables and the remaining ''irrelevant'' degrees of freedom, i.e.,
friction and correponding stochastic noise \cite{AB_00}, and possibly,
memory. 

\subsubsection{Dynamic rescaling}
\label{sec:multiscale_dynamics_rescaling}

Gaining a more quantitative understanding of the dynamic speedup
between FG and CG models is an interesting problem of
statistical mechanics. One promising approach is excess entropy
scaling. The idea goes back to the ''principle of corresponding
states'' as formulated by  Helfand and Rice in 1960 \cite{HR_60},
which states that, for fluids of particles interacting with a
potential of the form $V(r) = \epsilon u^*(r/\sigma)$, the dynamical
and transport properties for different $\sigma$ and $\epsilon$ can be
mapped onto each other.  For that particular choice of potential, the
correspondence can be shown by simple dimensional analysis and seems
close to trivial, but it does establish an interesting correlation
between dynamic and thermodynamic quantities. Building on this and a
method to map simple fluids onto hard sphere fluids
\cite{hansen_mcdonald_book}, Rosenfeld proposed a heuristic approach
to identifying corresponding states in simple fluids based on their
excess entropy \cite{Rosenfeld_77, Rosenfeld_99}, which turned out to
be remarkably successful both in the dense and dilute limit.
Recently, Rondina et al \cite{RBM_20} have shown that excess entropy
scaling can also be applied in dense polymer melts. They used a simple
bead spring model as a starting point which was coarse-grained to
different degrees and showed that the ratio of dynamical quantities
like the bond relaxation time $\tau$ or the viscosity $\eta$ in the CG
and FG system followed an exponential law 
\begin{equation}
X_{\mbox{\tiny CG}} \big/ X_{\mbox{\tiny FG}} = A \: \exp(\alpha \Delta
S_{\mbox{\tiny exc}})
\end{equation}
in a wide temperature range (with $X = \tau$ or $\eta$). Here $\Delta
S_{\mbox{\tiny exc}}$ is the temperature-dependent excess entropy
difference between the CG and the FG system which was determined by
thermodynamic integration.  However, the correspondence was found to be
less universal than one might hope, since both $\alpha$ and $A$ depend
on the CG model.

Lyubimov, Guenza and coworkers \cite{LCCG_10,LG_11,LG_13} have
considered CG models that map polymer melts onto a fluid of
interacting soft blob, and designed a first-principles approach that
allowed them to estimate the speedup factor with remarkable accuracy.
They assumed that the speedup factor has two contributions: The first
is based on a simple rescaling according to the principle of
corresponding states \cite{HR_60}. The second accounts for the
different environments of the interacting units, i.e., the different
effective friction constants of monomers that are part of a tagged
chain and of tagged CG particle in a melt environment. Both are
calculated within mode coupling theory \cite{Schweizer_89} and then
mapped onto each other. Using this Ansatz, Guenza et al were able to
derive analytical expressions for the dynamic speedup factor of
diffusion constants in chemically realistic melts such as
polybutadiene \cite{LG_13}. Unfortunately, the calculations require a
rather involved analytic machinery, and extensions to complex
inhomogeneous systems and mixtures are not yet available.

\subsubsection{Introducing friction}
\label{sec:multiscale_dynamics_friction}

In particle-based CG models, natural frameworks for introducing
friction are the Langevin thermostat, which allows to assign separate
friction constants for every CG particle, or the DPD thermostat, which
conserves momentum and allows to adjust independently the friction
parameters for every pair of interacting CG beads. 

A natural generic way to determine CG friction parameters from FG
simulations is provided by the Green-Kubo formalism, which relates
the friction force experienced by a particle moving at fixed velocity
to the integral over the time correlation function of the fluctuating
forces acting on the particle (the FACF). In the language of linear
response theory, this expression relates a steady-state generalized
''current'' (in this case the mean force on the particle), which
builds up in response to a constant ''thermodynamic force'' (in this
case the fixed velocity), to the time-integrated current-current (in
this case force-force) correlations.  Phrased in this way, one can
immediately see why a na\"ive application of the approach to FG
simulation trajectories is dangerous: The velocity of the CG particles
is not fixed, instead it fluctuates and averages to zero, and as a
result, the integral over the FACF vanishes as well \cite{AB_00}. If
the time scales of the dynamics of CG variables and remaining
irrelevant variables are well-separated, one can overcome this problem
by monitoring the running Green-Kubo integral as a function of an
upper time cutoff.  It will then first reach a more or less
well-defined plateau before it starts decaying, and the plateau value
can be used to extract values for the friction parameter
\cite{LBCK_14}.  However, the choice of the time cutoff value remains
somewhat heuristic. 

One way to overcome this so-called ``plateau problem'' is to constrain the
dynamics of the FG simulations such that the momentum of the CG
particles is kept fixed.  A corresponding bottom-up scheme for
determining DPD friction parameters from FG simulations was first
proposed by Akkerman and Briels \cite{AB_00} and later derived more
formally by Hij\'{o}n et al  \cite{HEVD_10} based on the Mori-Zwanzig
formalism and an additional Markovian assumption.  The idea is to
modify the dynamics of the FG simulations such that the desired
collective CG variables are constrained to fixed values and do not
participate in the dynamics. This effectively decouples the FG
dynamics from the CG dynamics and solves the plateau problem.
Hij\'{o}n et al \cite{HEVD_10} demonstrated the power of the approach
using the example of star polymer melts, and several other authors
have later applied it to derive CG DPD models for chemically realistic
oligomers or polymers such as $n$-alkanes \cite{TSPC_14},
polybutadiene \cite{LCR_17}, and dimethylpropane \cite{DV_18}. 

\subsubsection{Introducing memory}
\label{sec:multiscale_dynamics_memory}

The strategy of absorbing the full dynamics of the irrelevant variables
in a single set of DPD friction parameters is justified if the time scales 
in the CG model are well separated from those processes in the
FG system that have been integrated out \cite{AB_00}.  However, if the
degree of coarse-graining is comparatively low, or if the CG model does
not capture all slow processes in the FG model, the time scale
separation of characteristic processes at the FG and the CG level is
incomplete. In such cases, the Mori-Zwanzig projection technique
\cite{zwanzig_book} yields CG dynamical equations that are Non-Markovian, i.e.,
include memory terms.  Two types of approaches have been adopted in
the past to account for this effect. 

The first is to introduce virtual, but physically motivated variables
that mimic the effect of slow processes that have been eliminated in the
CG model \cite{NOB_07,Briels_09}, while not affecting the structural and
thermodynamic properties of the CG system.  One example are the slip
links discussed in Section \ref{sec:scales_blobs}, which are introduced
to restore the effect of entanglements -- i.e., the slow dynamics of
topological constraints that are removed in extremely coarse-grained
models. Wu et al have recently developed a systematic method to derive
slip link parameters from experiments or FG simulation data
\cite{WMM_21}. Another example is the RaPiD model for polymers developed
by Briels and coworkers, which uses the center of mass of molecules as
CG variables, but introduces a set of additional virtual variables that
characterize the conformational state of the
molecules \cite{LOB_14,AGB_18}.

The second approach is to cast the dynamical equations in the CG model
in the form of generalized Langevin equations (GLEs), i.e., to
explicitly include memory in the CG dynamical equations
\cite{Review_Memory_21}. Setting up such equations is a
highly non-trivial task. In particle-based CG models, one would
ideally like to use a multidimensional GLE of the type
\begin{equation}
\label{eq:gle}
M_i \dot{\mathbf{V}}_i(t)
 = \mathbf{F}_i^C(t) - \int_0^t \!\! \ud s \sum_j 
 \mathbf{K}_{ij}(t,s) \: \mathbf{V}_j(s)
 + \partial \mathbf{F}_i(t),
\end{equation}
where $M_i$ and $\mathbf{V}_i$ are the mass and velocity of
CG particles, $\mathbf{F}_I^C(t)$ and $\partial \mathbf{F}_i(t)$
the conservative and fluctuating stochastic forces acting on them, and
$\mathbf{K}_{ij}(t,s)$ is a multidimensional memory kernel that
depends on the configuration at time $t$ and $s$, and which is 
related to the stochastic force by a fluctuation-dissipation 
relation
\begin{eqnarray}
\langle  \partial \mathbf{F}_i(t) \partial \mathbf{F}_j(s)
\rangle & =& m_j \sum_k \mathbf{K}_{ik}(t,s) \:
\langle \mathbf{V}_k(s) \mathbf{V}_j(s) \rangle
\nonumber \\
& \overset{\mbox{\tiny equil.}}{=} &
k_B T  \: \mathbf{K}_{ij}(t,s).
\end{eqnarray}
Note that we have used a tensor notation here, and the last
equality $\overset{\mbox{\tiny equil.}}{=}$ uses the relation
$\langle \mathbf{V}_j(s) \mathbf{V}_k(s) \rangle = 
\mathds{1} \delta_{jk} m_k k_B T$, which is valid in 
thermodynamic equilibrium. The form (\ref{eq:gle}) has been derived by
Kinjo and Hyodo \cite{KH_07} based on Mori-Zwanzig projections and
additional approximations \cite{GS_21, Review_Schilling_22}.  The
fluctuation-dissipation relation can also be derived from the
Mori-Zwanzig projection operator formalism, but one can show that it
is a general feature of GLEs which satisfy an orthogonality condition
for the relation between random force and velocity \cite{JS_21}.
Several methods have been developed and analyzed that allow to
determine memory kernels from FG simulations and thus construct
GLE-based CG models in a bottom-up fashion \cite{Review_Memory_21,
SBB_87, SBR_90, SKTL_10, CVR_14, LBL_16, JHS_17, JHS_18, KDKS_19,
MPS_19, MWSS_21, WMP_20, BSJS_21, KV_21, Hanke_21, KV_22, VM_22}. 

The main practical problem with the CG equation (\ref{eq:gle}) is that
simulations of such high-dimensional coupled integro-differential
equations are computationally very time consuming, mostly due to the
high costs associated with the generation of multidimensionally
correlated noise that satisfies the fluctuation-dissipation relation.
Therefore, simplifications must be made. The simplest and most
efficient one is to ignore all cross-memory terms and replace
$\mathbf{K}_{ij}(t,s)$ by a single scalar function,
$\mathbf{K}_{ij}(t,s) = \mathds{1} \delta_{ij} K(t-s)$. This approach
has been used, among other, by Wang et al \cite{WLP_19, WMP_20} to model
polymers in dilute solution, and by Klippenstein et al to model
polymer melts \cite{KV_21}. In their approach, Klippenstein et al
explicitly address the issue of multibody correlations and propose a
method to replace them by a single effective self-memory kernel. To
this end, they introduce a new scheme for consistently including the
cross-correlations between the stochastic forces and the conservative
interactions with the effective medium, which turns out to be quite
accurate in their studies of star polymer melts \cite{KV_21} and
Asakura-Oosawa fluids \cite{KV_22}.

Going beyond pure self-memory kernels, Li and coworkers have suggested
an approach, termed ''Non-Markovian DPD''(NM-DPD), which decomposes
the memory kernel into a sum of frequency dependent DPD friction
functions \cite{LBLK_15, LBYK_16, LLDK_17}. This relieves the noise
generation problem, as the stochastic forces can then be decomposed in
the same pairwise manner.  Unfortunately, the approach puts severe
constraints on the self-memory part of the memory-kernel
$\mathbf{K}_{ii}$, since it assumes $\mathbf{K}_{ii} = - \sum_{j \neq
i} \mathbf{K}_{ij}$. This can cause problems, e.g., when
considering hydrodynamic interactions between CG particles in implicit
solvent \cite{JHS_18}, or diffusion of molecules in penetrant
networks \cite{DV_18}. To overcome the problem, Jung et al have
developed a more general scheme for reconstructing and treating pair
memory kernals that decouples self- and pair-friction while still
ensuring linear scaling for short-ranged pair-interactions
\cite{JHS_18}. 

When comparing the two approaches to account for memory in CG systems
-- physically motivated virtual variables and GLE-based CG models --
we should note that the practical solution of GLE equations also often
involves the use of auxiliary variables \cite{Review_Memory_21}.
However, these auxiliary variables are just a numerical trick to solve
the GLE and have no physical meaning \cite{Mori_65b, FG_79, MG_83,
CBP_09, CBP_10}.  The idea is to replace the GLE by a set of regular
Langevin equations in an extended phase space.  This is possible if
the memory kernel can be approximated by a finite sum of possibly
complex, but decaying exponentials (a Prony series) \cite{BB_13}.
Alternatively, the parameters of the Langevin equations can be
determined directly from correlation functions obtained in FG
simulations \cite{WMP_20, BSJS_21} in a numerically well-controlled
manner \cite{BSJS_21}.

\begin{figure*}[t]
\includegraphics[width=0.9\textwidth]{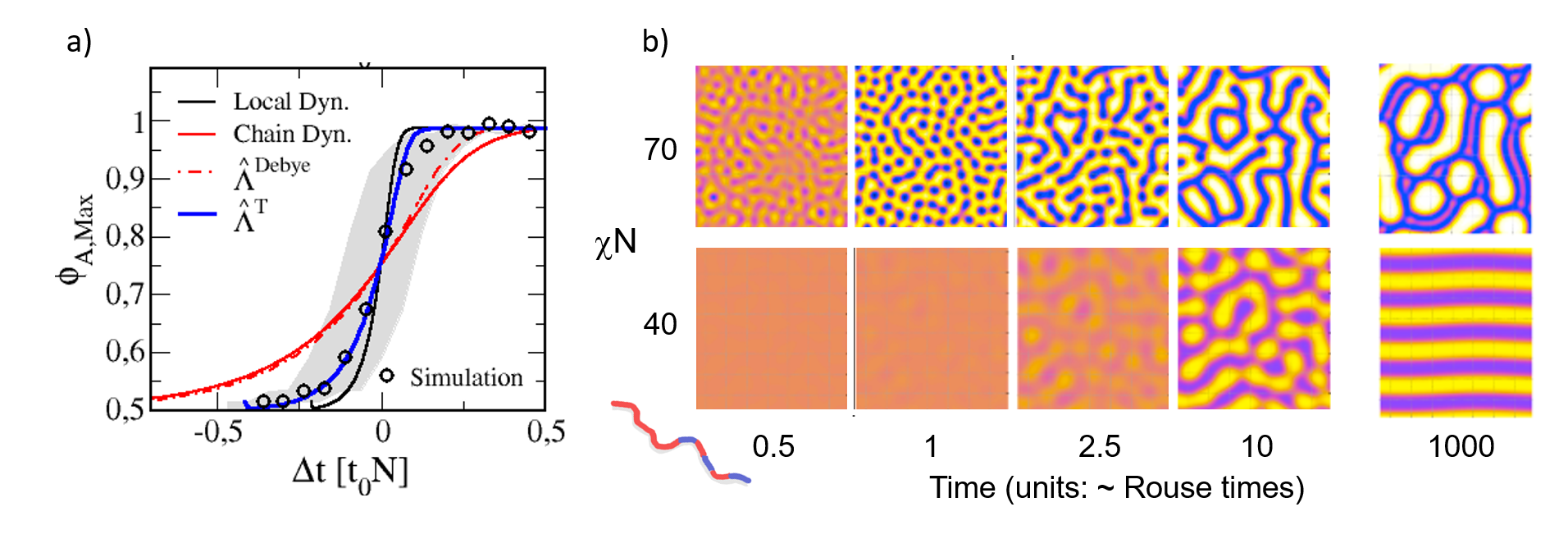} \caption{
a) Time evolution of order parameter in a diblock copolymer melt
after a sudden quench according to different DDFT models (line)
and CG particle-based simulations (symbols). The DDFT functional
$\hat{\Lambda}^T$ that has been contructed from the particle model.
Adapted from Ref.\ \protect\cite{MQS_20}.
b) Snapshots during the ordering of a melt of two-scale multiblock
copolymers ($A_{5n}B_nA_nB_nA_nB_n$ structure)  after a sudden
deep (top) and shallow (bottom) quench.
Adapted from Ref.\ \protect\cite{LDS_21}.
}
\label{fig:DDFT} 
\end{figure*}

\subsubsection{Transition particle-continuum}
\label{sec:multiscale_dynamics_particle-continuum}

So far, we have discussed dynamical coarse-graining issues in
particle-based models. Closely related problems arise in dynamical
coarse-graining from particle to continuum equations, when the CG are
dynamic equations for continuous fields. If the fields describe
complex fluids, e.g., polymer systems, one again needs to a account
for a multitude of time scales \cite{Guenza_99, PB_02, Panja_10, HK_16}
which often precludes the use of, e.g., simple Cahn-Hilliard type
equations \cite{Semenov_86}. Wang et al \cite{WRM_19} have considered
dynamic density functional (DDFT) equations of the type
(\ref{eq:ddft}), but with the mobility function replaced by a
time-delayed memory function
$\underline{\underline{\Lambda}}(\vec{r}-\vec{r}'; t-t')$ which they
calculated analytically in random phase approximation. They studied
the effect of memory on the ordering/disordering kinetics in
homopolymer and block copolymer melts and found very good agreement
between particle-based simulations and continuum
simulations \cite{WRM_19}.  One key to success in such an approach is
to identify the appropriate CG collective variables (densities).  Very
recently, M\"uller analyzed this problem \cite{Mueller_22} by examining
three situations where seemingly identical initial density
perturbations were created in different ways, first by applying a
modulated force on all segments of a melt, second by applying a force
on end segments only, and third by applying a force on a selected
middle segment. In particle-based simulations, the dynamic response to
these perturbations was found to be very different in the three cases.
This could be reproduced in the continuum simulations if the densities
of segments which had experienced the initial force and those of
passive segments were treated as separate collective variables.

From the point of view of dynamic coarse-graining, polymers have the
convenient feature that relaxation processes on different time scales
can often roughly be associated with different length scales.
Therefore, Markovian DDFT models such as (\ref{eq:ddft}) may be able
to capture the multiscale dynamics if the nonlocal mobility matrix
function is adjusted properly
$\underline{\underline{\Lambda}}(\vec{r}-\vec{r}')$.  Mantha et al
\cite{MQS_20} have developed a bottom-up method to construct
$\underline{\underline{\Lambda}}(\vec{r}-\vec{r}'; t-t')$ from
reference FG particle simulations, and found that simulations based on
the resulting DDFT model are in very good agreement with corresponding
particle-based simulations (see, e.g., Fig.\ \ref{fig:DDFT}).
Matching mobility matrices is also a convenient way to map different
CG particle-based polymer models onto each other \cite{LDS_21}. 

One should note, however, that these approaches are restricted to
systems close to equilibrium. Far from equilibrium, a CG continuum
description operating with densities only is certainly not sufficient
and one needs to introduce additional variables that characterize the
chain conformations \cite{MSS_20}. Furthermore, polymer stretching
generates mechanical stresses, therefore, the use of a purely
diffusive dynamical model such as (\ref{eq:ddft}) (model B dynamics)
is no longer justified. An appropriate CG model in such cases must
also include momentum and hydrodynamics.  As we have discussed in
Sec.\ \ref{sec:scales_macroscopic}, viscoelastic models are quite
commonly constructed from molecular polymer models such as the elastic
dumbbell model. Up to now, this is mostly based on analytical
considerations using substantial mean-field approximations, and to the
best of the author's knowledge, systematic bottom-up strategies to
construct full viscoelastic models from FG simulations are still
missing.  So far, coarse-graining strategies that connect particle
models with continuous viscoelastic models are mostly based on
parameter mapping \cite{BDEH_21}. 


Alternatively, it is sometimes possible to combine multiscale
approaches with theoretical insights, e.g., from the tube model of
viscoelasticity, in order to answer specific questions. One example of
such a theoretically informed scale bridging scheme is the
Branch-on-branch algorithm by McLeish and coworkers \cite{DIRK_06,
RADD_11}. It builds on a theory for the rheology of branched polymers,
the so-called Pom-Pom model by Bishko et al \cite{BMHL_97}, which
estimates the relaxation modulus of chains for given polymer
architectures, requiring only a fuew additional input parameters. The
Branch-on-branch algorithm establishes a connection between
microscopic simulations of the synthesis of highly branched polymers
and their rheological properties.  The microscopic simulations are
used to generate a representative sample of branched polymers, which
are then fed into the theoretical machinery. The method was applied
successfully by Read et al commercial Low density polyethylene (LDPE),
and recently by Zentel and coworkers to predict the rheological
properties of LDPE and polybutylacrylate (PBA) from the reaction
conditions in a miniplant \cite{ZDB_20, ZB_22, ZBPB_21, ZBDP_21,
GELZ_21}. 


\subsubsection{Accessing late times}
\label{sec:multiscale_dynamics_late-times}

So far, we have mainly discussed strategies to infer correct and
consistent dynamical properties from CG models. However, even though
CG simulations can cover much longer time spans than atomistic
simulations, this is usually  not sufficient to access experimentally
relevant time scales of seconds, minutes, minutes, or even months. 
In order to do so, one must also coarse-grain in time. 

In situations when it is possible to identify single activated events
that slow down the time evolution of a system, one can resort to rare
event sampling techniques. Typical problems that are considered with
such approaches are, e.g., refolding events of molecules or nucleation
processes in materials. A large portfolio of methods has been proposed
to study them \cite{YSZY_19}, such as weighted ensemble techniques
\cite{HK_96, ZJZ_10}, transition path sampling \cite{DBCC_98, BCDG_02,
BS_21}, Forward Flux Sampling \cite{AWT_05, RVT_09, BS_10}, the string
method \cite{ERV_02, ERV_05}, or combinations thereof \cite{AG_13}.
Using such methods, one can extract rate constants that can be fed
into a kinetic model in order to simulate a system on larger time
scales.

More generally, one of the most powerful frameworks for coarse
graining in time is Markov State Modelling, which has become very
popular in the field of biomolecular simulations \cite{PBB_10,
MSM_book_14, SS_15, HP_18}. In Markov State Models (MSMs), the
configurational space is partitioned into many regions, called
macrostates, and the dynamics is modelled in terms of a master
equation by memory-less transitions between these states. 
The number of macrostates is typically chosen quite large,
much larger than, e.g., the number of known metastable configurations
of a system. Replacing the original molecular dynamics equation by
such a relatively fine-grained Markovian process thus represents a
severe approximation. In reality, the transitions between macrostates
usually have some memory of the past. Nevertheless, it can be shown
that for optimized mappings, the long-term dynamics is still
reproduced very accurately by the MSM, as long as it is governed by a
few dominant slow time scales \cite{SS_15}. 

Nowadays, Markov State Modelling is a well-established technique with
solid theoretical foundations \cite{SS_15}. Techniques are available
how to optimize the partitioning into macrostates \cite{PPGD_13}, how
to determine transition rates and the associated uncertainty
\cite{TWPN_15}, and even, how to use MSMs to connect theoretical
models to experimental trajectory data \cite{KPN_12}. Recent Machine
learning based approaches offer additional opportunities for further
optimization \cite{MN_21, ALSN_22}.  Many of these techniques assume
equilibrium, i.e., they require transition rates to fulfill detailed
balance. Knoch and Speck have recently considered nonequilibrium MSMs
(NE-MSM)s for driven systems \cite{KS_15, KS_17, KS_19} and shown how
to connect MSMs at different CG levels (i.e., with different microstate
numbers) in a thermodynamically consistent manner. Knoch et al also
applied the MSM approach to the non-equilibrium problem of
force-driven molecule unfolding and showed that MSMs can be used to
bridge between loading rates in simulations to experimentally
accessible loading rates \cite{KS_18}. 

\begin{figure*}[t]
\includegraphics[width=0.9\textwidth]{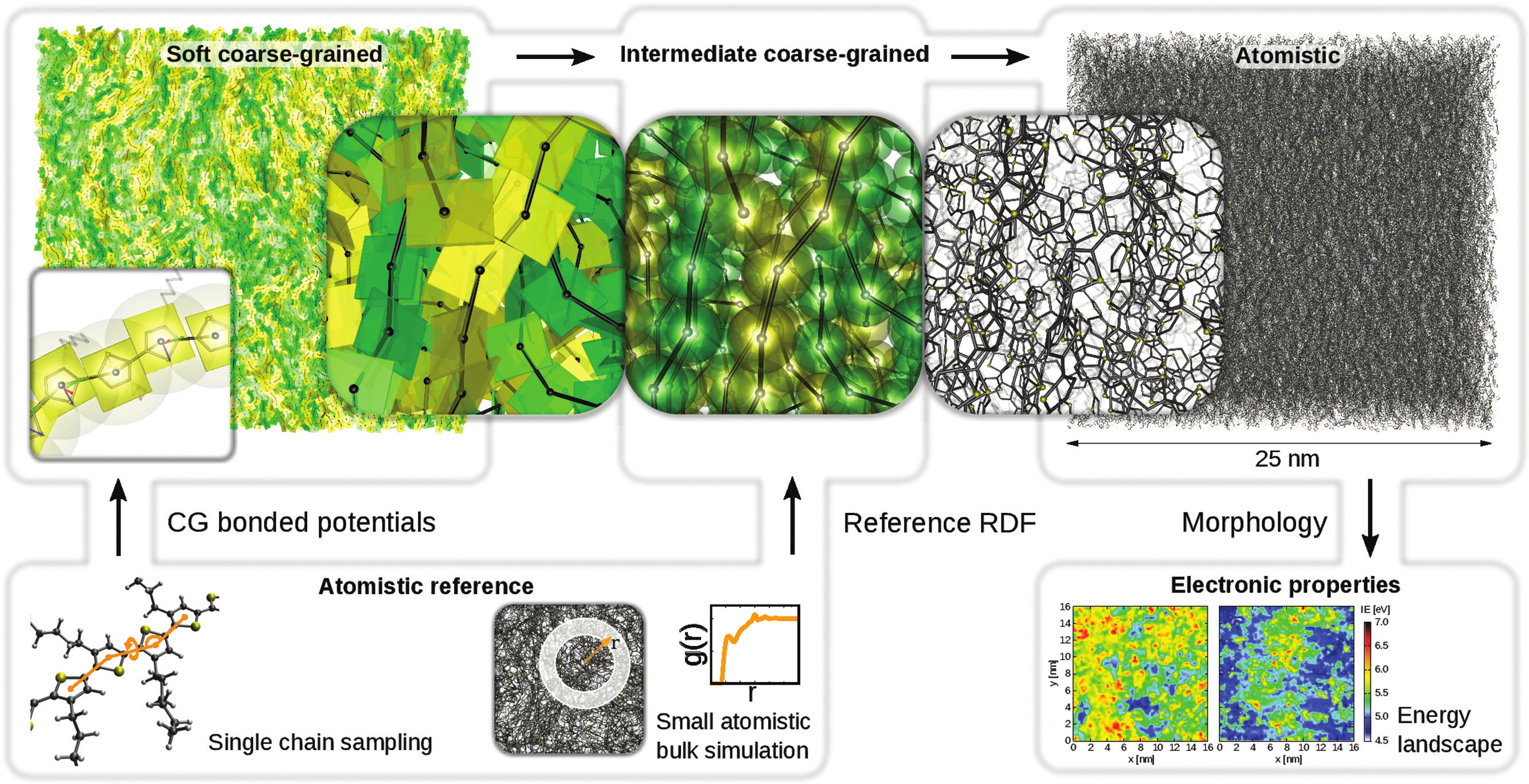}
\caption{Multiscale strategy for predicting charge transport in
polymeric semiconductors \protect\cite{GPKD_15}. See text for explanation.  
Reprinted from Reference \protect\cite{GPKD_15} with permission of XXX.}
\label{fig:semiconductors} 
\end{figure*}

\subsection{Multiresolution}

The ultimate vision of multiscale modelling is to study the properties
of a system simultaneously on different scales. Often, it is sufficient
to establish one-way communication channels between simulations at
different CG level.  As an example for such a scheme, Fig.\
\ref{fig:semiconductors} shows a strategy to predict the electronic
transport properties of polymeric organic materials, which has been
developed by Andrienko, and coworkers\cite{GPKD_15, KVAK_15, GMKA_19}.
It relies on a theoretical framework that relates the charge transport
in organic semiconducting polymers to their local atomistic
conformations\cite{AKMF_07, KMNK_07, NKKF_09, RLMS_11}, based on the
Marcus theory of electron transfer rates \cite{Marcus_93}. In the
multiscale approach of Andrienko and coworkers, coarse-grained and
ultra-coarse grained force fields for conjugated polymers are
constructed from atomistic reference simulations. Ultra-coarse grained
simulations based on these force-fields are then used to sample large
scale morphologies, which serve as starting point to create atomistic
configurations by a successive backmapping strategy (see below).  From
the atomistic structures, a local ionization energy landscape is
constructed, which allows to infer electronic properties such as the
charge mobility.

This work flow demonstrates the power of multiscale approaches,
however, it does not yet allow to account for possible feedback
mechanisms between processes on different scales. One way to
include them is provided by the ''heterogeneous multiscale'' (HMM)
framework proposed by E and Engquist in 2003 \cite{EE_03}. The idea is
to couple a macroscopic continuum simulation -- in their case a fluid
dynamics simulations -- with microscopic FG simulations, which serve
to estimate missing data for the macroscale model on the fly
\cite{RE_05,EELR_07,BLR_13}. This approach has recently been applied
by Lukacova-Medvidova and coworkers to study non-Newtonian flows of
shear-thinning polymers melts in complex geometries \cite{SYES_18,
DYSK_21, TGYL_21}. The HMM idea can also be extended to other types of
continuum models. For example, Honda and Kawakatsu \cite{HK_07} and
M\"uller and Daoulas \cite{MD_11} have proposed related
mixed-resolution models that concurrently couple time-dependent
Ginzburg Landau (TGL) models of (co)polymer blends to more detailed
models of the same system: The long-time evolution of the composition
profiles is simulated by TGL simulations, but SCF calculations
\cite{HK_07} or particle-based simulations \cite{MD_11} are carried
out intermittently, using the current TGL conformation as a basis, to
re-adjust the parameters of the TGL model.

In some situations, it may be desirable to study large portions of a
system with a coarse-grained model, but be able to zoom into selected
regions in space with higher resolution. This concept goes back to the
famous quantum mechanical/molecular mechanics (QM/MM) method
\cite{WL_76} by Warshel and Levitt, which combines electronic
structure calculations in selected regions of space with classical
atomistic molecular dynamics in the rest of the system, separated by a
hybrid transition region.  In a similar spirit, Prapotnik, delle Site
and Kremer in 2005 \cite{PDK_05} and de Fabritiis, Delgado-Buscalioni
and Coveney in 2006 \cite{DDC_06} have proposed mixed resolution
dynamical simulation schemes for complex fluids that allow zooming
into selected regions of space and studying them at higher resolution
-- the ''AdResS'' scheme \cite{PDK_05} and the ''hybrid MD'' scheme
\cite{DDC_06}. Both, however, differ from QM/MM approaches or related
approaches involving a CG outer model \cite{NACM_05} in one important
aspect: They allow for a particle exchange between the CG and FG
regions. The AdResS scheme \cite{PDK_05} achieves this by implementing
a gradual switch between FG and CG force-fields in a transition region
-- or, in a later ''Hamiltonian-AdResS'' variant \cite{PFED_13}, a
switch between interaction potentials -- and the hybrid-MD scheme
\cite{DDC_06} couples a particle model to a continuum model via flux
boundary conditions and allows to generate and remove particles in the
transition region. The approaches have subsequently been refined and
extended  in various directions \cite{DF_07,ENMK_07, OGD_21}, combined
with each other \cite{DKP_08,DKP_09}, and related schemes have been
proposed.  For example, Qi et al have developed a scheme that couples
field-based and particle-based polymer models \cite{QBS_13, QBRS_16,
QS_17b} via a spatially varying semi-grandcanonical potential that
enforces identity switches \cite{Abrams_05}. As an interesting new
idea for AdResS, Heidari et al \cite{HKCP_18, HKGP_20, BDSH_21}
recently proposed to use an ideal gas as outer medium. While the
latter hardly qualifies as a high-level model of a complex soft
material, the setup allows to carry out simulations with true open
boundaries, to determine absolute free energies for the coupled FG
system, and to enforce nonequilibrium situations with, e.g.,
steady-state current \cite{HKGP_20}.

\begin{figure*}[t]
\includegraphics[width=0.95\textwidth]{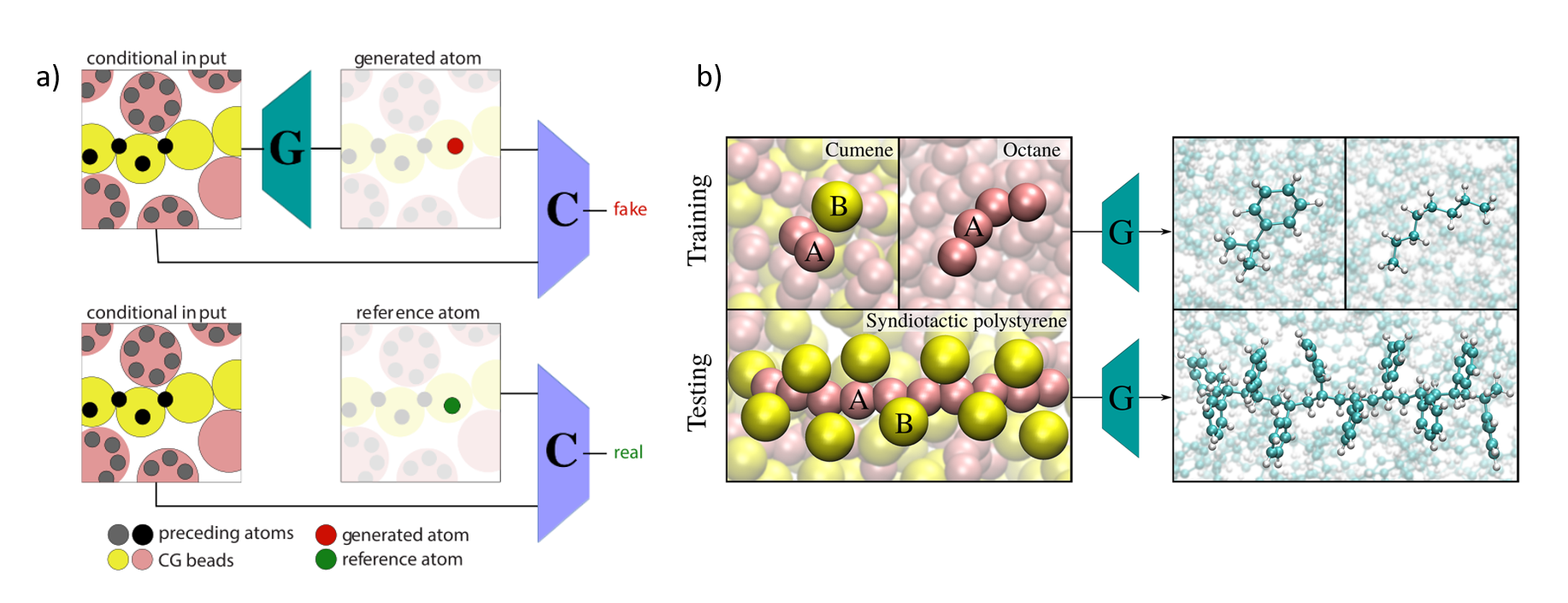}
\caption{Backmapping strategy for polymer melts based on generative 
adversarial networks (GANs). (a) Sketch of the approach: A generator
network $G$ sequentially samples atom positions depending on the 
CG structure and the existing atoms. The discriminator $C$ evaluates
true and fake configurations based on the discrepancy between
reference atoms and generated atoms. The training objective of $G$ 
is to fool $C$, and the training objective of $C$ is not to be fooled.
Reprinted from Ref.\ \protect\cite{SWB_20}  with permission of XXX.
(b) A GAN trained on cumene and octane can be used for backmapping
of syndiotactic polystyrene.
Reprinted from Ref.\ \protect\cite{SBW_21}  with permission of XXX.
}
\label{fig:GAN}
\end{figure*}

A crucial component of many multiresolution schemes is backmapping: To
connect different levels of resolution to each other, one must not
only go up the scales by coarse-graining, but also be able to go down,
i.e., generate representative microscopic molecular conformations from
CG configurations. This is often done in a two-step procedure: First,
some heuristic method is used to construct a first guess for the
positions of the FG particles, and then, the FG configuration is
further optimized by simulated annnealing or another energy
minimization method \cite{TKHB_98,HAVK_06, SMMM_07, RSGR_10, LMC_16}.
Liu et al have proposed an alternative method based on configurational
bias Monte Carlo \cite{LSLV_08}. The backmapping problem is
particularly challenging in the case of proteins, due to their complex
chemical structure, and a number of sophisticated methods have been
developed for this community. An overview can be found in Ref.\
\cite{KGKW_16}, Table 3. More recently, machine learning tools
borrowed from computer graphics become increasingly attractive. The
reason is that the backmapping problem has some similarity with
typical problems in computer graphics, such as, filling a given rough
frame with a representative set of textures.  Stieffenhofer et al
\cite{SWB_20, SBW_21} and Li et al \cite{LBPH_20} have recently
developed backmapping methods based on deep generative adversarial
networks (GANs), a framework where two neural networks -- a
''generator'' and a ''discriminator'' compete with each other in order
to learn the main statistical properties of a training set in an
unsupervised manner.  Stieffenhofer et al tested their scheme on
syndiotactic polystyrene and showed that it can create good backmaps
already before energy minimization,  and that it is transferable:
A GAN trained on melts can be used for backmapping of polymer crystals
and even chemically similar (small) molecules \cite{SWB_20, SBW_21}
(see Fig.\ \ref{fig:GAN}). 


An interesting application of multiresolution tools is the generation
of equilibrated polymer melt configurations for large molecular weight
-- which is still a difficult problem in polymer
simulations \cite{SKHF_16}. In traditional approaches, one first
prepares a reasonably random initial configuration, e.g., by
assembling a number of polymers with typical melt configurations, and
then further relaxes it by implementing unphysical dynamics and/or
Monte Carlo moves that allow chain crossing or even change chain
connectivity \cite{AEGK_03, MZMS_15, OMGM_15,SKC_16}. In multiscale
approaches \cite{Subramanian_10, Subramanian_11, NKM_14, NMGA_20,
ZMSD_14, ZSDK_15, OKD_18, ZCHS_19}, one uses CG simulations to
equilibrate the melt and then reconstructs a FG configuration by
increasing the level of resolution in a stepwise fashion. Tubiana et
al have recently performed a systematic comparison of a traditional
and multiscale equlibration scheme, focussing on topological
indicators such as knot distributions \cite{MHV_18}, and found
excellent agreement \cite{TKPD_21}. 

\bigskip

%

\subsection{Machine-Learning based strategies}
\label{sec:multiscale_machine-learning}

Regarding virtually all aspects of scale bridging techniques discussed
above, machine-learning (ML) based methods are becoming increasingly
important \cite{Review_NTMC_20}. Kernel-based techniques or artificial
neural networks (ANNs) are used for identifying suitable CG variables
\cite{VBR_20, ALSN_22}, for determining accurate atomistic potentials
that bridge between ab-initio calculations and standard classical
force fields \cite{Review_UCSG_21, MMKU_21} as well as for deriving
improved coarse-grained potentials  \cite{BAL_15, JC_17, GZV_18,
SSAB_20, BN_21, Review_Webb_21}, for determining memory kernels from
FG simulations \cite{WMP_20}, for constructing MSMs \cite{MN_21,
ALSN_22}, or for backmapping \cite{LBPH_20, SWB_20, SBW_21}. In some
cases, ANNs can be trained to predict the outcome of CG simulations -
such as, e.g., the conformation of heteropolymers for a given monomer
sequence \cite{WJGP_20}, the equation of state of homopolymer melts
for given pair potential \cite{BN_21}, or even complex quantities such
as drug-membrane insertion free energies\cite{HMKB_19}. This opens
interesting perspectives for new coarse-graining strategies or new
strategies of materials design.

Traditionally, an important application of ML in polymer science has
been to predict material properties of interest of
polymer based materials or composites, such as, e.g., the tribological
properties, wear resistance, thermal conductivity \cite{ZF_03, RWHB_19,
LVR_21, GERS_22, NTL_22}, or even self-assembly  \cite{THLL_20}.
Specifically, ANNs have been used for some time to predict the glass
transition temperature $T_g$, using as input either the chemical
structure only \cite{MJ_02, Ning_09, Liu_10, LTZ_12}, or additional
information from small-scale quantum mechanical
calculations \cite{LC_09, PEF_12}.  Depending on the materials, the
predictive power can be quite high \cite{HHMY_19, PILM_19, MS_20a,
MS_20b, WLWL_20, TVL_21, KAXR_21, CTL_21, ZX_21}. Such approaches can
be applied, e.g., for identifying promising candidates for ''high
temperature polymers''\cite{TCL_21} with high $T_g$. More generally,
a central vision in the emerging field of polymer
informatics \cite{AP_17, WJGP_20, Bereau_21, GW_21, JJWS_21} is to
provide ML tools for the discovery of new interesting
polymer materials.  Ramprasad and coworkers have recently launched a
polymer informatics platform (www.polymergenome.org) which offers
tools for predicting a variety of polymer properties that include the
density, $T_g$, the melting temperature, the dielectric constant, the
tensile strength, and many others \cite{KCTD_18,TKCC_20}.

One should note that, in general, extracting new physical insights
from ML-based scale bridging strategies is not a trivial task.  They
can help to unveil hidden structure-property relations or correlations
between different material properties \cite{MS_20a, MLZL_19}, but they
do not necessarily explain the underlying mechanisms. When used for
coarse-graining and force field generation, they can be viewed as
being a sophisticated interpolation scheme between available
(training) data, but they do not necessarily help to understand
general principles of coarse-graining or general features of CG
models. On the other hand, feeding in theoretical knowledge of
physical principles enhances the efficiency of ML-based procedures and
reduces the amount of necessary training data \cite{Review_NTMC_20}.
The resulting ML-based models usually have much higher predictive
power than the original purely knowledge-based models.  Hence
knowledge-based and ML-based approaches to scale bridging and
multiscale modelling mutually fertilize each other and should be seen
as complementary.

\section{Challenges for the Future} \label{sec:challenges}
Comparing the available scale-bridging techniques in the last
section to the examples of scale-bridging phenomena in polymers in the
introduction, it becomes clear that we still need to go a long way
before these two ends meet.  Being able to gain a comprehensive
quantitative understanding of real-world phenomena that includes the
interplay of structures processes on all relevant scales, from the
smallest to the largest, remains a grand challenge of polymer science.
Several hard problems still need to be solved. 

\begin{description}

\item{\em Strong inhomogeneities}. Real polymeric
materials are usually  multiphase materials, they are filled with
internal interfaces and their composition varies dramatically in
space. Therefore, the transferability issues that still afflict most
CG models represent serious problems that need to be overcome, e.g.,
by further improving coarse-graining strategies or by
linking different CG models to each other in a multiresolution
sense. 

\item{\em Defects}. Defects are omnipresent in soft materials. They
can be defined as very dilute and strongly localized perturbations
that may come in several flavors \cite{Review_Defects_20}: As doping
defects in the form of local impurities, as connectivity defects that
distort the local molecule structure, or as topological defects that
do not involve any local chemical modifications, but still locally
perturb the structure in a manner that they cannot be removed without
global rearrangements of the whole system (e.g., dislocations).
Because they are highly dilute, they are usually not present in small
scale simulations unless forced to be there; neverheless, they tend to
have a large and long-range impact on the material properties. In
order to study this, small scale simulations should thus capture the
effect of a defect that they actually do not contain, and that imparts
its presence only, e.g., via non-periodic boundary conditions.

\item{\em Nonequilibrium and Processing}. As discussed above, many
traditional CG concepts are developed for equilibrium systems or at
least build on a local equilibrium assumption. On the other hand,
already the example of viscoelastic phase separation shows that in
polymers, local equilibrium cannot be taken for granted even in
seemingly simple problems such as spinodal phase separation. Most
polymeric materials never reach equilibrium and their properties
crucially depend on the way they have been created \cite{CBCF_19}.
Therefore, quantitative multiscale descriptions must be able to
account for processing pathways. The practical importance of
nonequilibrium processes in polymer systems has been acknowledged for
a long time, and nonequilibrium phenomena as occur, e.g., in polymer
rheology, have been a research focus since the beginnings of polymer
science. Nevertheless, systematic scale bridging strategies for
nonequilibrium polymers are still in their infancy.

\item{\em Accessing late times}.  The time bridging strategies
discussed in Sec.~\ref{sec:multiscale_dynamics_late-times} have mostly
been applied to single molecules or simple small systems, and not to
materials. In order to understand the properties of polymeric
materials at late times, depending on environmental conditions, and
phenomena such as ageing, abrasion and wear, failure and fatigue, one
must account for all factors listed above (inhomogeneities, defects,
processing history) and study their (co-)evolution over a very long time
period. So far, theoretical models \cite{VYSN_18, KKRB_22, SGKR_22}
are mostly based on empirical theories and with little connection to
microscopic simulations.

\end{description}

Multiscale modelling of polymers thus remains a difficult problem, but
it also offers exciting new prospects for the future. For example, one
fascinating challenge will be to develop systematic multiscale
strategies for truly nonequilibrium living polymeric systems as are
common in biology, which depend on strongly fluctuating local
compositions and are constantly driven out of equilibrium. 

\section*{Acknowledgment} 

This work is funded by the German Science Foundation within the
Collaborative Research Center TRR 146 ''multiscale simulation methods
for soft matter systems'' via Grant 233630050.  The author has had the
privilege to serva as the spokesperson of this center for many years.
She has benefitted from many stimulating discussions and enjoyable
interactions with all members of this Konsortium, in particular
Tristan Bereau, Kostas Daoulas, Gregor Diezemann, Burkhard D\"unweg,
Martin Hanke, Kurt Kremer, Arash Nikoubashman, Joe Rudzinski, Thomas
Speck, Lukas Stelzl, Nico van der Vegt, Peter Virnau, and Michael
Wand.


\bibliographystyle{achemso}
\bibliography{polymers}


\end{document}